\documentclass[review,12pt]{elsarticle}

\usepackage{geometry}
\usepackage{amssymb}

\usepackage{lineno}
\usepackage{amsmath,amsthm,amsfonts}
\usepackage{url}
\usepackage{tikz}

\biboptions{sort&compress}

\usepackage{subcaption}

\newcommand{\ud}{\mathrm{d}}

\newcommand{\Bi}{\text{Bi}}
\newcommand{\erf}{\text{erf}}

\newcommand{\be}{\begin{equation}}
\newcommand{\ee}{\end{equation}}
\newcommand{\bea}{\begin{eqnarray}}
\newcommand{\eea}{\end{eqnarray}}
\newcommand{\beas}{\begin{eqnarray*}}
\newcommand{\eeas}{\end{eqnarray*}}
\newcommand{\bi}{\begin{itemize}}
\newcommand{\ei}{\end{itemize}}

\newcommand\nc{\newcommand}
\nc{\Oc}{\mathcal{O}} \nc{\Omo}{\Omega_{\tiny{\mbox{out}}}}
\nc{\Omi}{\Omega_{\tiny{\mbox{in}}}}
\nc{\Omit}{\Omega_{\tiny{\mbox{int}}}}
\nc{\Pio}{\Pi_{\tiny{\mbox{out}}}}
\nc{\Pii}{\Pi_{\tiny{\mbox{in}}}}
\nc{\Piit}{\Pi_{\tiny{\mbox{int}}}}
\nc\pa{\partial}
\nc\pad[2]{\frac{\pa #1}{\pa #2}} 
\nc\padd[2]{\frac{\pa^2 #1}{\pa{#2}^2}}
\nc\nd[2]{\frac{\ud #1}{\ud #2}}
\nc\ndd[2]{\frac{d^2 #1}{d {#2}^2}}
\nc\pat[2]{\frac{D #1}{D#2}} 
\nc\ov{\overline} 
\nc\degree{^{\circ}} 
\nc\ord[1]{{\cal O}(#1)} 
\nc\ra{\rightarrow} 
\nc\Ra{\Rightarrow} 
\nc\dint{{\mbox ~d}}
\nc\dg{{\dot \gamma}}
\nc\units[1]{$^\text{#1}$}

\begin{document}

\begin{frontmatter}



\title{Non-Local Effects and Size-Dependent Properties in Stefan Problems with Newton Cooling}%

\author[Tim,Tim2]{Marc Calvo-Schwarzw\"alder\corref{cor1}}

\cortext[cor1]{mcalvo@crm.cat}
\address[Tim]{Centre de Recerca Matem\`{a}tica, Campus de Bellaterra, Edifici C, 08193 Bellaterra, Barcelona, Spain.}
\address[Tim2]{Departament de Matem\`{a}tiques, Universitat Polit\`{e}cnica de Catalunya, 08028, Barcelona, Spain.}

\begin{abstract}
We model the growth of a one-dimensional solid by considering a modified Fourier law with a size-dependent effective thermal conductivity and a Newton cooling condition at the interface between the solid and the cold environment. In the limit of a large Biot number, this condition becomes the commonly used fixed-temperature condition. It is shown that in practice the size of this non-dimensional number is very small. We study the effect of a small Biot number on the solidification process with numerical and asymptotic solution methods. The study indicates that non-local effects become less important as the Biot number decreases. 
\end{abstract}


\end{frontmatter}


\section{Introduction}\label{sec:intro}
It is widely accepted that heat conduction at small length scales differs from the classical description \cite{Chang2008,Cahill2003,Font2018b}. There exist a wide range of theoretical models which extend Fourier's law to account for non-local effects which become dominant on length scales comparable to the phonon mean free path (MFP). The models may be classified into micro-, meso- or macroscopic models depending on whether the aim is focused on describing the behaviour of the individual heat carriers, the evolution of their distribution or related macroscopic quantities, such as the temperature or the heat flux. Examples for micro- and mesoscopic models are molecular dynamics, Monte Carlo simulations \cite{Chen2005}, the Boltzmann transport equation (BTE) \cite{Boltzmann1872} or the equation of phonon radiative transfer \cite{Majumdar1993}. The equations involved in these approaches often require excessive computational effort and it is therefore more sensible to aim for a macroscopic equation which captures the physics of the problem. Popular macroscopic models are the Maxwell--Cattaneo law \cite{Cattaneo1958}, the thermomass model \cite{Wang2010,Dong2011} or the Guyer--Krumhansl (GK) equation \cite{Guyer1966a,Guyer1966b} and the framework of phonon hydrodynamics developed from it \cite{Dong2011,Jou1996,Guo2015,Torres2018}. 

Based on the BTE and on the framework of extended irreversible thermodynamics \cite{Jou1996}, Alvarez and Jou \cite{Alvarez2007} derived a model for heat flow with a size-dependent effective thermal conductivity (ETC). This was able to capture non-local effects and showed good agreement with experimental data. Hennessy et al. \cite{Hennessy2018} show that such an ETC can also be derived when the GK equation is included into the formulation of the Stefan problem. Recently, Font \cite{Font2018} proposed a simpler non-classical formulation of the Stefan problem which includes an effective Fourier law where a size-dependent ETC replaces the bulk thermal conductivity. He showed that, even when the temperature at the boundary is instantly set to the temperature of the cold environment, the problem of an initially infinite solidification rate is avoided by incorporating a size-dependent ETC.

The popular fixed-temperature condition is a specific limit of a more general boundary condition known as the Newton cooling condition. In this paper we aim to generalise the model proposed by Font by considering cooling conditions at the interface between the material and its environment. For large Biot (or Nusselt) numbers, this condition converges to the fixed-temperature condition. Formulations of the Stefan problem where Newton cooling conditions are applied, can be found in the framework of nanoparticle melting \cite{Ribera2016,Myers2016,Hennessy2018b} or nanowire melting \cite{Florio2016}. Hennessy et al. \cite{Hennessy2018b} recently proposed an extended form of cooling condition accounting for memory effects. However, the case of small Biot number is not considered in these studies. In Sec. \ref{ssec:params} we explain the importance of this limit.


\section{Mathematical model}\label{sec:model}
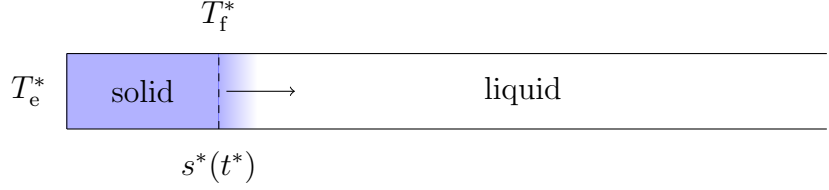
\begin{figure}[h!]
\centering
\begin{tikzpicture}
	\shade[left color=blue,right color=blue,opacity=0.3] (0,0) rectangle (2,1);
	\shade[left color=blue,right color=white,opacity=0.3] (2,0) rectangle (2.5,1);
	\draw (0,0) -- (10,0);
	\draw (0,1) -- (10,1);
	\draw (0,0) -- (0,1);
	\draw [dashed](2,0) -- (2,1);
	\draw [->] (2.1,.5) -- (3,.5);
	\node at (1,.5) {solid};
	\node at (6,.5) {liquid};
	\node at (2,-.5) {$s^*(t^*)$};
	\node at (2,1.5) {$T^*_\text f$};
	\node at (-.5,.5) {$T^*_\text e$};
\end{tikzpicture}
\caption{A liquid bath, initially at the phase change temperature $T^*_\text f$, starts to solidify due to a low external temperature $T^*_\text e$.}
\label{fig:scheme}
\end{figure}

We consider a liquid bath, initially at the phase change temperature $T^*_\text f$, that occupies the space $x^*\geq0$. The $^*$ notation refers to dimensional quantities. Due to a low external temperature $T^*_\text{e}$ applied at $x^*=0$, the liquid starts to undergo a solidification process and a solid starts to grow into the liquid occupying the space $0\leq x^*\leq s^*(t^*)$; see Fig. \ref{fig:scheme}. For simplicity, the specific heat $c^*$, the density $\rho^*$ and the bulk thermal conductivity $k^*$ are assumed to be constant, and no supercooling will be considered.

\subsection{Classical formulation}
The process is driven by the heat flow in the solid phase. The liquid remains at the phase change temperature $T^*_\text f$ for all time. The temperature $T^*$ and heat flux $q^*$ in the solid are related via conservation of energy, which in one spatial dimension takes the form
\begin{equation}\label{cons_energy}
	c^*\rho^*\pad{T^*}{t^*}=-\pad{q^*}{x^*}.
\end{equation} 
In the classical formulation, the heat flux is determined by the gradient of the temperature,
\begin{equation}\label{Fourier}
	q^*=-k^*\pad{T^*}{x^*}.
\end{equation}
At the solid-liquid interface, the temperature of the solid is equal to the phase change temperature,
\begin{equation}\label{model:bcs}
	T^*(s^*,t^*)=T^*_\text f.
\end{equation}
The heat exchange at the interface between the material and the environment is governed by a Newton cooling condition
\begin{equation}\label{model:bc0}
	q^*(0,t^*)=h^*(T^*_\text e-T^*(0,t^*)),
\end{equation}
where $h^*$ is the heat transfer coefficient. Contrary to the fixed temperature condition $T^*=T^*_\text e$, which corresponds to the limit $h^*\to\infty$, the cooling condition does not lead to an infinite initial solidification rate \cite{Ribera2016,Florio2016,Gupta2003,Alexiades1992}.

The evolution of the interface is determined by an energy balance which is known as the Stefan condition In its simplest form, this is
\begin{equation}\label{model:Stefan}
	\rho^* L^*_m \nd{s^*}{t^*}=-q^*(s^*,t^*),
\end{equation}
where $L^*_m$ is the latent heat of fusion. 

Finally, at $t^*=0$ we assume that no solid has been produced,
\begin{equation}
	s^*(0)=0.
\end{equation}

\subsection{Effective Fourier Law}\label{ssec:modelEIT}
For a system of characteristic size $L^*$, Alvarez and Jou \cite{Alvarez2007} propose substituting the bulk thermal conductivity $k^*$ by an ETC of the form
\begin{equation}\label{model:keff1}
	k^*_\text{eff}(L^*)=2k^* \left(\frac{L^*}{\ell^*}\right)^2\left(\sqrt{1+\left(\frac{\ell^*}{L^*}\right)^2}-1\right).
\end{equation}
where $\ell^*$ is the phonon mean free path. The ratio $\ell^*/L^*$ is the Knudsen number and it determines the dominance of non-local effects in heat transport. In the limit $L^*\ll\ell^*$ (Kn$\gg1$), Eq. \eqref{model:keff1} reduces to $k^*\approx 2k^*_0L^*/\ell^*$, which is in accordance with experimental data \cite{Li2003}. Conversely, in the limit $\ell^*\gg L^*$ (Kn$\ll1$) the classical limit $k^*_\text{eff}=k^*$ is retrieved. In the current problem the system length-scale is determined by the size of the solid region, hence $k^*_\text{eff}(L^*)=k^*_\text{eff}(s^*)$ and we obtain
\begin{equation}\label{EIT:Fouriermod}
	q^*=-k^*_\text{eff}(s^*)\pad{T^*}{x^*},
\end{equation}
which we assume to govern heat conduction through the solid. Upon combining Eqs. \eqref{cons_energy} and \eqref{EIT:Fouriermod} we obtain 
\begin{equation}
	c^*\rho^*\pad{T^*}{t^*}=k^*_\text{eff}(s^*)\pad{^2T^*}{x^{*2}},
\end{equation}

The temperature at the solid-liquid interface is determined by Eq. \eqref{model:bcs}, whereas at $x^*=0$ Eq. \eqref{model:bc0} becomes
\begin{equation}\label{EIT:bc0}
	-k^*_\text{eff}(s^*)\pad{T^*}{x^*}\bigg|_{x^*=0}=h^*(T^*_\text e-T^*(0,t^*)).
\end{equation}
The Stefan condition may now be written as
\begin{equation}\label{EIT:Stefan}
	\rho^* L^*_m \nd{s^*}{t^*}=k^*_\text{eff}(s^*)\pad{T^*}{x^*}\bigg|_{x^*=s^*}.
\end{equation}

\subsection{Dimensionless formulation}

The temperature scale is given by the temperature jump at $x^*=0$, that is, $\Delta T^*=T^*_\text f-T^*_\text e$. The natural length scale of the problem is the mean free path $\ell^*$ and hence we choose the corresponding diffusive time scale $c^*\rho^*\ell^{*2}/k^*$, which balances the terms in Eq. \eqref{cons_energy}. Consequently, we introduce the dimensionless quantities $x=x^*/\ell^*$, $s=s^*/\ell^*$, $t=k^*t^*/(c^*\rho^*\ell^{*2})$ and $T=(T^*-T^*_\text f)/\Delta T^*$. Upon writing derivatives as indices, the dimensionless equations are
\begin{subequations}\label{nd:model}
\begin{alignat}{3}
	&T_t=f(s)T_{xx},\qquad &&0<x<s,\label{nd:heat}\\
	&f(s) T_x=\Bi(1+T),\qquad &&x=0,\label{nd:bc0}\\
	&T=0,\qquad &&x=s,\label{nd:bcs}\\
	&\beta s_t=f(s)T_x,\qquad &&x=s,\label{nd:Stefan}\\
	&s=0,\qquad &&t=0,\label{nd:ic}
\end{alignat}
\end{subequations}
where $\beta=L^*_m/(c^*\Delta T)$ and $\Bi=h^*\ell^*/k^*$ are the Stefan number and the Biot number respectively. In melting problems, the latter is also called Nusselt number because the material in contact with the environment is in liquid state. Finally, $f(s)=2s\left(\sqrt{s^2+1}-s\right)$ is the non-dimensional form of $k^*_\text{eff}$.

Note, due to the way in which the problem has been scaled, the mean free path has been eliminated from the ETC and is only present in the Biot number. In fact, the dimensionless position of the interface $s(t)$ can now be understood as an effective Knudsen number depending on time, since it corresponds to the ratio of the size of the growing solid to the mean free path.

The effect of the Stefan number on the evolution of the interface in formulations with a fixed temperature has been studied by many authors; see the books by Gupta \cite{Gupta2003} or Alexiades and Solomon  \cite{Alexiades1992}, for example. However, when Newton cooling conditions are applied, the behaviour of the system when varying Bi has to be studied as well.

\subsection{Parameter estimation}\label{ssec:params}
%

In this paper we will use the thermophysical parameters of silicon, since it is material that is widely used in theoretical studies \cite{Alvarez2007,Font2018,Li2003,Font2017,Torres2017b}. The values of the relevant physical quantities are given in Table \ref{table:silicon}.

\begin{table}
	\centering
	\caption{Thermophysical parameters for Silicon at 1000 K \cite{Torres2017b,Mills2000}.}\label{table:silicon}
	\begin{tabular}{cccccc}
		\hline
		\hline
			$k^*$ [W/m$\cdot$K] & $c^*$ [J/kg$\cdot$K] & $\rho^*$ [kg/m$^\text{3}$] & $L^*_m$ [kJ/kg] & $T^*_\text{f}$ [K] & $\ell^*$ [$\mu$m]\\
		\hline
			43.67 & 864.89 & 2296 & 1787 & 1687 & 12.84\\
		\hline
		\hline
	\end{tabular}
\end{table}

The Stefan number can be parametrized in terms of the temperature change as $\beta = \mathcal{T}^*/\Delta T$ and where $\mathcal{T}^*=L^*_m/c^*$. Using the values given in  Table \ref{table:silicon} we find $\mathcal{T}^*\approx2754$ K. Therefore, even for a temperature drop of 100 K we still obtain $\beta\approx27.5$. For other materials such has tin, lead or gold, we find that $\mathcal{T}^*$ is of the order of hundreds of Kelvin \cite{Ribera2016,Hennessy2018b,Font2013}, and hence we can still expect large Stefan numbers for temperature changes of tens of Kelvin. However, since the liquid is assumed to be initially at the phase change temperature, only a small temperature change is needed to drive the solidification process and thus we expect $\beta\gg1$. 

The Biot number can be expressed as $\Bi=h^*/\mathcal{H}^*$, where for silicon $\mathcal{H}^*=k^*/\ell^*\approx3.4\times10^{9}$ W/m$^2\cdot$K. Determining the heat transfer coefficient $h^*$ is complicated, since its value depends on the environment and the material. Nonetheless, there is a maximal value $h^*_\text{max}$ beyond which the material would simply vaporise. Ribera and Myers \cite{Ribera2016} provide the expression $h_\text{max}^*=c^*\sqrt{\rho^*B^*/3}$, where $B^*$ is the bulk modulus. For silicon we have $B^*\approx10^{11}$ kg/m$\cdot$s$^2$ \cite{AzomSilicon}, therefore $h^*_\text{max}\approx7.6\times10^{9}$ W/m$^\text{2}\cdot$K and hence $\Bi_\text{max}\approx2.2$. However, the order of magnitude of $h^*_\text{max}$ is extremely large and is never reached in practical situations \cite{Bamberger1986}, thus $h^*\ll h^*_\text{max}$ generally.
 
Hence, throughout this study we will assume $\beta\gg1$ and $\Bi\ll1$. If necessary, the relative size of $\beta$ to Bi will be discussed during the analysis.

\section{Solution methods}\label{sec:solmethods}
There exist few analytical solutions to practical Stefan problems. However, in the case of a constant thermal conductivity, i.e. $f(s)=1$, and a fixed temperature condition at $x=0$, it is possible to obtain an exact solution, termed the Neumann solution \cite{Gupta2003,Alexiades1992,Hill1987}.
 
Such a solution does not exist with a Newton cooling condition, not even in the case of constant thermal conductivity, and thus we need to explore numerical and approximate methods to solve the problem with $f\neq1$. Furthermore, the expected sizes of the Biot and Stefan numbers allow us to perform an asymptotic analysis to distinguish different time regimes of interest and obtain analytical expressions in most of them.

\subsection{Numerical solution}
A usual approach to obtain numerical solutions of Stefan problems is to first introduce an alternative variable which transforms the moving domain into the unit interval. The numerical scheme to solve the resulting problem consists of discretizing explicitly for the temperature and implicitly for the interface position and the interface speed \cite{Font2018,Ribera2016,Font2013,Font2015}. In addition, since the solid does not exist initially, a small time analysis must be performed to obtain a valid initial condition for the numerical scheme.

\subsubsection{Boundary fixing transformation}
We define the new variable $\xi=x/s$ and rewrite the temperature as $T(x,t)=u(\xi,t)$, which transforms Eqs. \eqref{nd:heat}--\eqref{nd:Stefan} into
\begin{subequations}\label{numerics:eqsfixed}
\begin{alignat}{3}
	&su_t=\xi s_tu_\xi+F(s)u_{\xi\xi},\qquad &&0\leq \xi\leq 1,\label{numerics:heat}\\
	&F(s)u_\xi=\Bi(1+u),\qquad &&\xi=0,\label{numerics:bc0}\\
	&u=0,\qquad &&\xi=1,\label{numerics:bc1}\\
	&\beta s_t=F(s)u_\xi,\qquad &&\xi=1,\label{numerics:Stefan}
\end{alignat}
\end{subequations}
where we have defined $F(s)=2(\sqrt{1+s^2}-s)$ for simplicity. This formulation breaks down as $t\to0$ and hence the initial condition for $s$ must be substituted by an approximate value at a small time $t_0$.

\subsubsection{Small-time solution}\label{ssec:smalltime}
Since the solid phase does not exist initially, a small-time analysis has to be carried out to study the initial dynamics of the system. 
To facilitate the analysis, we assume the remaining dimensionless numbers are order one in magnitude. In this way, the small-time behaviour applies to more limits that may arise in other situations where, for instance, the Stefan number is expected to be small \cite{Alexiades1993}. An analysis of the results for asymptotic limits of these parameters can be performed afterwards. Let now $t\ll1$ and assume that the solid-liquid interface is approximated by an expression of the form $s\approx\lambda t^p$ (hence $s_{t}\approx p\lambda t^{p-1}$), where $\lambda$ and $p$ are constants to be determined. Neglecting terms of order $t^{2p}$ gives $F\approx2(1-\lambda t^p)$ and thus Eq. \eqref{numerics:Stefan} can be written as
\begin{equation}\label{smalltime:Stefan}
	\frac{1}{2}\beta \lambda p\left(t^{p-1}+\lambda t^{2p-1}\right) =u_\xi,
\end{equation} 
where we have used $(1-z)^{-1}\approx1+z$ for $z\ll1$ and $u_\xi$ is evaluated at $\xi=1$.
The only possible way to balance with the r.h.s. of Eq. \eqref{smalltime:Stefan} is obtained by setting $p=1$, hence 
\begin{equation}\label{smalltime:s}
	s\approx\lambda t,\qquad s_{t}\approx \lambda, \qquad F\approx2(1-\lambda t),
\end{equation}
provided $t\ll1$. In particular, Eq. \eqref{smalltime:s} predicts that the solid initially grows at a finite rate $\lambda$ independent of the choice of boundary condition at $x=0$. The fixed boundary condition is known to predict an infinite phase change rate \cite{Ribera2016,Florio2016,Gupta2003,Alexiades1992,Font2013,Font2015}, so the size-dependent ETC produces a more realistic growth rate.

Substituting Eq. \eqref{smalltime:s} into Eqs. \eqref{numerics:heat}--\eqref{numerics:bc1} and  taking the limit $t\to0$ yields a second order boundary value problem for $u$
\begin{subequations}\label{smalltime:ODEu}
\begin{align}
	&\lambda\xi u_\xi+2 u_{\xi\xi}=0,\\
	&2u_\xi(0)=\Bi(1+u(0)),\\
	&u(1)=0.
\end{align}
\end{subequations}
The solution to \eqref{smalltime:ODEu} is
\begin{equation}\label{smalltime:u}
	u(\xi)=\Bi\frac{\erf\left(\sqrt{\lambda}\xi/2\right)-\erf\left(\sqrt{\lambda}/2\right)}{2\sqrt{\lambda/\pi}+\Bi\,\erf\left(\sqrt{\lambda}/2\right)},
\end{equation}
where $\erf(z)=2\pi^{-1/2}\int_0^z\exp(-t^2)dt$ is the error function. Finally, $\lambda$ is determined by substituting Eqs. \eqref{smalltime:s} and \eqref{smalltime:u} into Eq. \eqref{smalltime:Stefan} and taking the limit $t\to0$, which yields an equation for $\lambda$
\begin{equation}\label{smalltime:lambda_Newton}
	\lambda+\frac{\sqrt{\pi}}{2}\Bi\sqrt{\lambda}\erf\left(\frac{\sqrt{\lambda}}{2}\right)=\frac{\Bi}{\beta}e^{-\lambda/4}.
\end{equation}
For a fixed value of $\Bi$ it is trivial to find $\beta(\lambda)$, which can be used to plot the exact solution on the $(\beta,\lambda)$-plane. The same argument is valid for $\Bi(\lambda)$ for a fixed $\beta$. However, finding $\lambda$ for fixed values of $\Bi$ and $\beta$ requires the use of numerical methods. 

Recall, Eq. \eqref{smalltime:lambda_Newton} has been obtained under the assumption $\Bi,\beta=O(1)$. Since we are interested in the limits $\Bi,\beta^{-1}\ll1$, we can perform an asymptotic analysis on Eq. \eqref{smalltime:lambda_Newton} to find an approximate solution in this specific limit.

\begin{figure}[h!]
	\centering
	\begin{subfigure}{.47\textwidth}
    \centering
    \includegraphics[width=\textwidth]{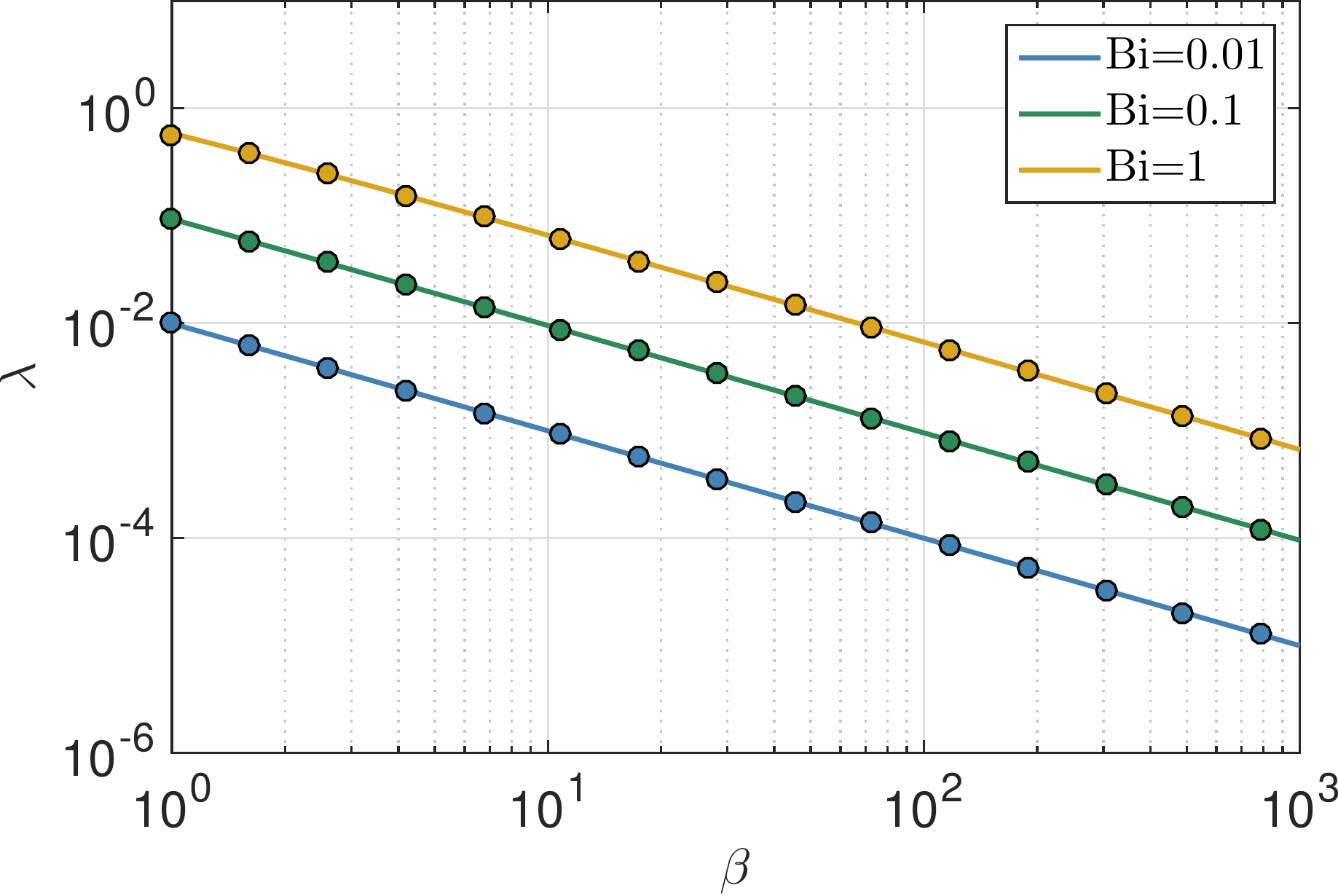}
    \caption{}\label{fig:lambda_Nu_a}	
	\end{subfigure}
	~
	\begin{subfigure}{.47\textwidth}
    \centering
    \includegraphics[width=\textwidth]{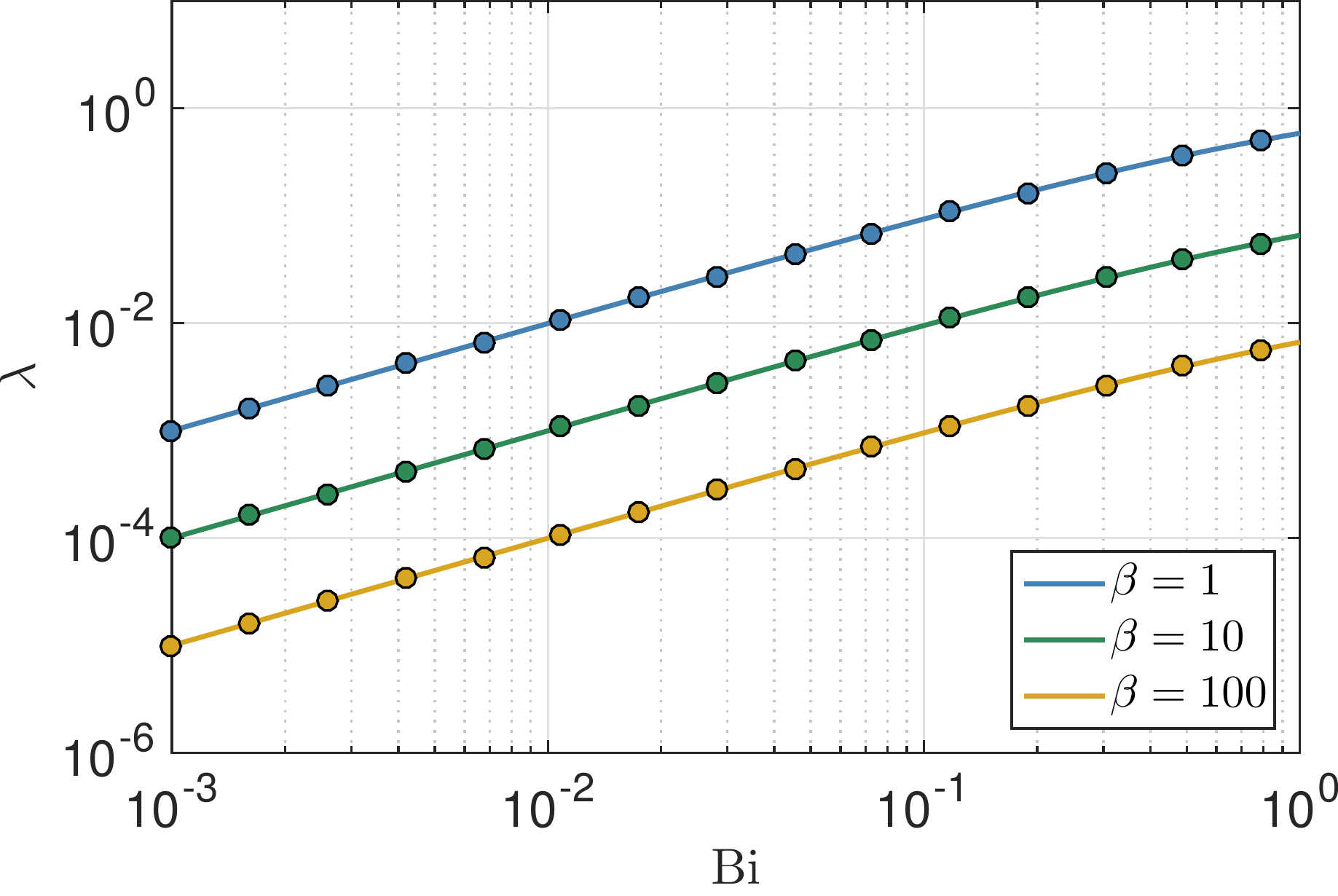}
    \caption{}\label{fig:lambda_Nu_b}	
	\end{subfigure}
	\caption{(a) Solution to Eq. \eqref{smalltime:lambda_Newton} as a function of $\beta$, for various values of the Biot number Bi. (b) Solution to Eq. \eqref{smalltime:lambda_Newton} as a function of Bi, for various values of the Stefan number $\beta$. Solid lines correspond to the exact solutions and circles refer to the asymptotic approximation given in Eq. \eqref{smalltime:asymptotics_lambda}.}
	\label{fig:lambda_Nu}	
\end{figure}

In the limit $\Bi,\Bi/\beta\to0$, Eq. \eqref{smalltime:lambda_Newton} yields $\lambda\to0$, which indicates $\lambda=O(\delta)$ for some $\delta=\delta(\Bi,\beta)\ll1$. We can make the approximations
\begin{equation}\label{smalltime:approximations}
	\erf\left(\frac{\sqrt{\lambda}}{2}\right)\approx\frac{\sqrt{\lambda}}{\sqrt{\pi}}-\frac{\lambda\sqrt{\lambda}}{12\sqrt{\pi}},\qquad \exp\left(\frac{\lambda}{4}\right)\approx1+\frac{\lambda}{4},
\end{equation}
which reduces Eq. \eqref{smalltime:lambda_Newton} to
\begin{equation}\label{smalltime:lambda_Newton_asy}
	\left(1+\frac{\Bi}{2}\right)\lambda+\frac{1}{4}\left(1+\frac{\Bi}{3}\right)\lambda^2+O(\lambda^3)=\frac{\Bi}{\beta}.
\end{equation}
Note, the r.h.s of Eq. \eqref{smalltime:lambda_Newton_asy} represents the driving force of the solidification process, therefore a balance with the l.h.s is required. Since $\lambda\ll1$, this balance is only achieved if we choose $\lambda=O(\Bi/\beta)$. Then we seek a solution for $\lambda$ as a power series
\begin{equation}
	\lambda=\delta\lambda_0+\delta^2\lambda_1+O(\delta^3).
\end{equation}
where $\delta=\Bi/\beta$. Introducing this expansion into Eq. \eqref{smalltime:lambda_Newton} yields a subproblem for each power of $\delta$. The first- and second-order subproblems are
\begin{equation}
	\left(1+\frac{\Bi}{2}\right)\lambda_0=1,\qquad \left(1+\frac{\Bi}{2}\right)\lambda_1+\frac{1}{4}\left(1+\frac{\Bi}{3}\right)\lambda_0^2=0,
\end{equation}
which gives $\lambda_0=2/(2+\Bi)$ and $\lambda_1=-(3+\Bi)\lambda_0^3/12$ and hence
\begin{equation}\label{smalltime:asymptotics_lambda}
	\lambda=\frac{2}{2+\Bi}\frac{\Bi}{\beta}-\frac{2(\Bi+3)}{3(2+\Bi)^3}\frac{\Bi^2}{\beta^2}+O\left(\frac{\Bi^3}{\beta^3}\right).
\end{equation}
In Fig. \ref{fig:lambda_Nu} we compare this approximation to the exact solution of Eq. \eqref{smalltime:lambda_Newton} and observe an excellent agreement across the entire range $\Bi,\beta^{-1}\in(10^{-3},1)$.

Consider now the first-order approximation. Substituting into Eq. \eqref{smalltime:u} and using Eq. \eqref{smalltime:approximations} we can reduce $u$ to
\begin{equation}\label{smalltime:asymptotics_u}
	u(\xi)\approx\frac{\Bi(\xi-1)}{2+\Bi},
\end{equation}
which may be used to approximate the temperature profile at a small time $t_0$ for $\Bi,\beta^{-1}\ll1$.

\subsubsection{Numerical scheme}\label{ssec:numerics}
The problem in the fixed domain is now discretized on a spatial grid of $N+1$ points of the form $\xi_i=\frac{i}{N}$ and a time grid with points of the form $t_n=t_0+n\Delta t$. In interior points of the domain, derivatives are approximated using second order central differences:
\begin{equation}
	\pad{u}{t}\approx\frac{u_{i}^{n+1}-u_{i}^{n}}{\Delta t},\quad \pad{u}{\xi}\approx\frac{u_{i+1}^{n+1}-u_{i-1}^{n+1}}{2\Delta\xi},\quad \pad{^2u}{\xi^2}\approx\frac{u_{i+1}^{n+1}-2u_{i}^{n+1}+u_{i-1}^{n+1}}{(\Delta\xi)^2},
\end{equation}
where $u^n_i=u(\xi_i,t_n)$ and $\Delta\xi=N^{-1}$. Upon substituting these expression into Eq. \eqref{numerics:heat} and writing $s=s^n$ and $\ud s/\ud t=s_{t}^n$ we obtain $N-1$ algebraic equations of the form
\begin{equation}
	A_{i}^n u^{n+1}_{i-1} + B_{i}^n u^{n+1}_{i} + C_{i}^n u^{n+1}_{i+1} = s^n u^{n}_{i},\qquad 1\leq i \leq N-1,
\end{equation}
with coefficients
\begin{equation}
	A_i^n =\frac{\xi_is_t^n\Delta t}{2\Delta\xi} -	\frac{F(s^n)\Delta t}{(\Delta\xi)^2},\quad
	B_i^n =s^n+2\frac{F(s^n)\Delta t}{(\Delta\xi)^2},\quad
	C_i^n =-\frac{\xi_is_t^n\Delta t}{2\Delta\xi} -	\frac{F(s^n)\Delta t}{(\Delta\xi)^2}.
\end{equation}
Using Eq. \eqref{numerics:bc1} we obtain
\begin{equation}
	u^{n+1}_N=0,
\end{equation}
whereas the Newton condition at $\xi=0$ becomes, after using second order forward differences,
\begin{equation}
	\left(1+\frac{3}{2}\alpha^n \right)u^{n+1}_0-2\alpha^nu^{n+1}_1+\frac{1}{2}\alpha^nu^{n+1}_2=-1,
\end{equation}
where $\alpha^n=F(s^n)/(\Bi\Delta\xi)$.
To update the position of the moving boundary we discretize Eq. \eqref{numerics:Stefan} explicitly for $s$ to find $s^{n+1}$,
\begin{equation}
	s^{n+1}=s^n+\Delta t F(s^n)\frac{3u^{n+1}_N-4u^{n+1}_{N-1}+u^{n+1}_{N-2}}{2\beta\Delta \xi},
\end{equation}
where we have used second order backward differences to discretize $u_\xi$ at $\xi=1$. To avoid stability issues, we discretize time as a logarithmically spaced grid. In this way, the time steps are smaller at the beginning of the process, where the growth rate is expected to be larger. Numerical experimentation shows that no unstabilities appear for the considered range of parameters.

\subsection{Asymptotic solution}\label{ssec:asy}

In \cite{Font2018}, an asymptotic analysis for the formulation with a fixed-temperature condition is performed under the assumption of a large Stefan number. In the case of the Newton cooling condition, the asymptotic analysis depends also on $\Bi$ and potentially on the relative size of Bi to $\beta$. In order to simplify the analysis we will focus only on three time regimes, which correspond to $s\ll1$, $s=O(1)$ and $s\gg1$.

The first time regime is given by $t=O(\varepsilon)$, where $\varepsilon\ll1$ is an artificial parameter, and it has already been partially studied during the small-time analysis. It describes the initial stage of the solidification, where the solid has not grown much yet, therefore $x,s=O(\epsilon_1)$ for some small $\epsilon_1\ll1$. In addition, the small Biot number indicates that the influx of heat is very small and therefore the temperature is not expected to differ much from its initial value, $T=O(\epsilon_2)$ for $\epsilon_2\ll1$. Upon defining the scaled variables $t=\varepsilon\hat t$, $x=\epsilon_1 \hat x$, $s=\epsilon_1 \hat s$ (thus $f\approx2\epsilon_1\hat s$) and $T=\epsilon_2 \hat T$, by balancing terms in the Newton condition we find $\epsilon_2=\Bi$. To ensure that solidification occurs at $\hat t>0$ we need to balance both sides of Eq. \eqref{nd:Stefan}, which yields $\epsilon_1=\Bi\beta^{-1}\varepsilon\ll\varepsilon$. At the leading order, the system in the new variables reads
\begin{subequations}
\begin{alignat}{3}
	&\hat T_{\hat x\hat x}=0,\qquad &&0\leq \hat x\leq \hat s,\label{asy:heat}\\
	&2\hat s\hat T_{\hat x}=1,\qquad &&\hat x=0,\label{asy:bc0}\\
	&\hat T=0,\qquad &&\hat x=\hat s,\label{asy:bcs}\\
	&\hat s_{\hat t}=2\hat s\hat T_{\hat x},\qquad &&\hat x=\hat s,\label{asy:Stefan}\\
	&\hat s=0,\qquad &&\hat t=0,\label{asy:ic}
\end{alignat}
\end{subequations}
After applying the boundary conditions \eqref{asy:bc0} and \eqref{asy:bcs}, the solution to Eq. \eqref{asy:heat} is
\begin{equation}
	\hat T=\frac{\hat x-\hat s}{2\hat s},
\end{equation}
and therefore the position of the interface is determined by
\begin{equation}
	\hat s=\hat t.
\end{equation}
In the original non-dimensional variables we find
\begin{equation}\label{asy:small_time}
	T(x,t)=\Bi\frac{x-s}{2s},\qquad  s(t)=\Bi\beta^{-1}t
\end{equation}
for $t\ll1$. Notice that this temperature profile coincides with the small-time profile given in Eq. \eqref{smalltime:asymptotics_u} when terms of order $\Bi^2$ are neglected.

The second time regime of interest is when $s=O(1)$, which invalidates the approximation $f\approx2s$. From Eq. \eqref{asy:small_time} we find $t=O(\Bi^{-1}\beta)$ and $T=O(\Bi)$. Upon defining the new variables $x=x^\prime$, $s=s^\prime$, $t=\Bi^{-1}\beta t^\prime$ and $T=\Bi T^\prime$, the leading order problem for $T^\prime$ becomes
\begin{subequations}
\begin{alignat}{3}
	&T^\prime_{x^{\prime}x^{\prime}}=0,\qquad &&0\leq x^\prime\leq s^\prime,\label{asy:heat2}\\
	&f(s^\prime)T^\prime_{x^\prime}=1,\qquad &&x^\prime=0,\label{asy:bc02}\\
	&T^\prime=0,\qquad &&x^\prime=s^\prime,\label{asy:bcs2}
\end{alignat}
\end{subequations}
whose solution is
\begin{equation}
	T^\prime(x^\prime,t^\prime)=\frac{x^\prime-s^\prime}{f(s^\prime)},
\end{equation}
The solid-liquid interface is therefore determined by
\begin{equation}
	s^\prime_{t^\prime}=1,\label{asy:Stefan2}
\end{equation}
subject to the matching condition
\begin{equation}
	s^\prime\sim t^\prime,\quad t^\prime\to0.
\end{equation}
In the original dimensional variables the solution in this time regime is therefore
\begin{equation}\label{asy:sol2}
	T(x,t)=\Bi\frac{x-s}{f(s)},\qquad s(t)=\Bi\beta^{-1}t,
\end{equation}
for $t=O(\Bi^{-1}\beta)$, which captures the previous time regime as well due to the behaviour of $f$ for small values of $s$.

The third time regime of interest is when $s\gg1$ and therefore $f\approx1$. The correct balance is obtained by choosing $t=O(\Bi^{-2}\beta)$, $x,s=O(\Bi^{-1})$ and $T=O(1)$, which yields, upon writing $t=\Bi^{-2}\beta\tilde t$, $x=\Bi^{-1}\tilde x$, $s=\Bi^{-1}\tilde s$ and $T=\tilde T$,
\begin{subequations}\label{asy:3}
\begin{alignat}{3}
	&\tilde T_{\tilde x\tilde x}=0,\qquad &&0\leq \tilde x\leq \tilde s,\label{asy:heat3}\\
	&f(\tilde s)\tilde T_{\tilde x}=1+\tilde T,\qquad &&\tilde x=0,\label{asy:bc03}\\
	&\tilde T=0,\qquad &&\tilde x=\tilde s,\label{asy:bcs3}\\
	&\tilde s_{\tilde t}=f(\tilde s)\tilde T_{\tilde x},\qquad &&\tilde x=\tilde s.\label{asy:Stefan3}
\end{alignat}
\end{subequations}
The solution to \eqref{asy:3} is given by
\begin{equation}
	\tilde T(\tilde x,\tilde t)=\frac{\tilde x-\tilde s}{\tilde s+ f(\tilde s)},
\end{equation}
and thus the interface is determined by
\begin{equation}
	\tilde s_{\tilde t}=\frac{f(\tilde s)}{\tilde s+f(\tilde s)},
\end{equation}
which can be integrated to give
\begin{equation}\label{asy:s_eq}
	C+4 \tilde t=\tilde s^2+\tilde s\sqrt{1+\tilde s^2}+4\tilde s+\text{arcsinh}(\tilde s),
\end{equation}
where $C$ is a constant of integration to be determined to match the previous time regime. Using Eq. \eqref{asy:sol2} we find $\tilde s\sim\tilde t$ for $\tilde t=O(\Bi)$, which gives $C=0$ at leading order.
In the original dimensionless variables the temperature profile is
\begin{subequations}
\begin{equation}
	T(x,t)=\frac{x-s}{s+\Bi^{-1}f(s)},
\end{equation}
whereas the solid-liquid interface is determined by
\begin{equation}\label{asy:s_eq_dim}
	\frac{4\Bi}{\beta}t=\Bi s^2+s\sqrt{1+(\Bi s)^2}+4s+\Bi^{-1}\text{arcsinh}(\Bi s).
\end{equation}
\end{subequations}
To obtain $s$ we must invert Eq. \eqref{asy:s_eq_dim} numerically. Equivalently, it can be calculated integrating 
\begin{equation}\label{asy:finalODE}
	s_t=\frac{f(s)}{\beta(s+\Bi^{-1} f(s))},
\end{equation}
which captures all the previous time regimes and can thus be solved using the initial condition $s(0)=0$.

We can consider a fourth time regime, where the Newton condition converges to the fixed temperature condition, as it can be observed by assuming $\tilde x\gg1$ in Eq. \eqref{asy:bc03}. However, observe that Eq. \eqref{asy:finalODE} already captures this regime, since $s\to\infty$ yields $f\to1$ and $s_t\propto s^{-1}$, from where the classical behaviour $s\sim\sqrt{t}$ is recovered.

\section{Results and discussion}\label{sec:results}
%

\begin{figure}[h!]
	\begin{subfigure}{.47\textwidth}
    \centering
    \includegraphics[width=\textwidth]{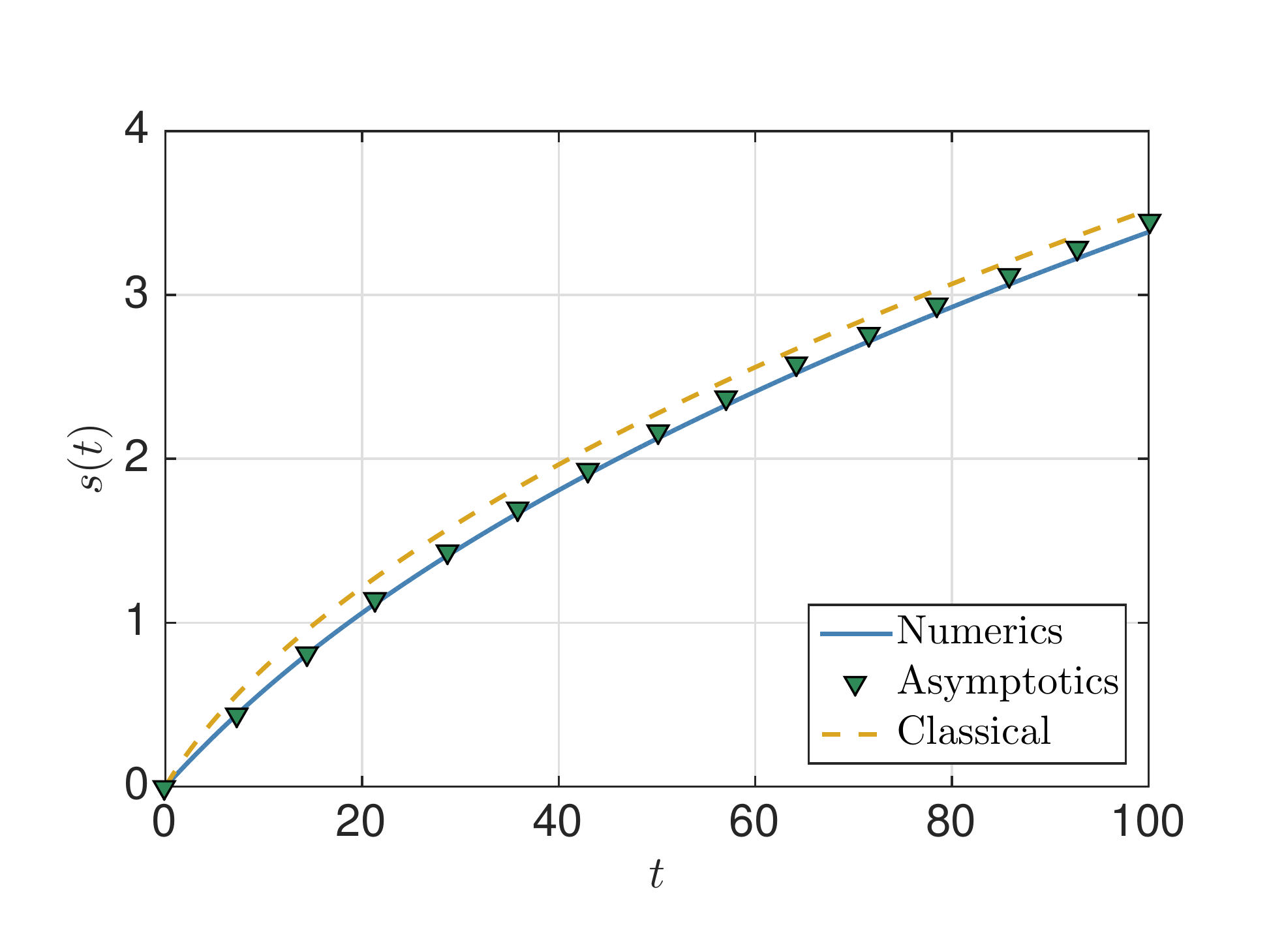}
    \caption{$\beta=10$, $\Bi=1$.}\label{fig:results_a}
	\end{subfigure}
	~
	\begin{subfigure}{.47\textwidth}
    \centering
    \includegraphics[width=\textwidth]{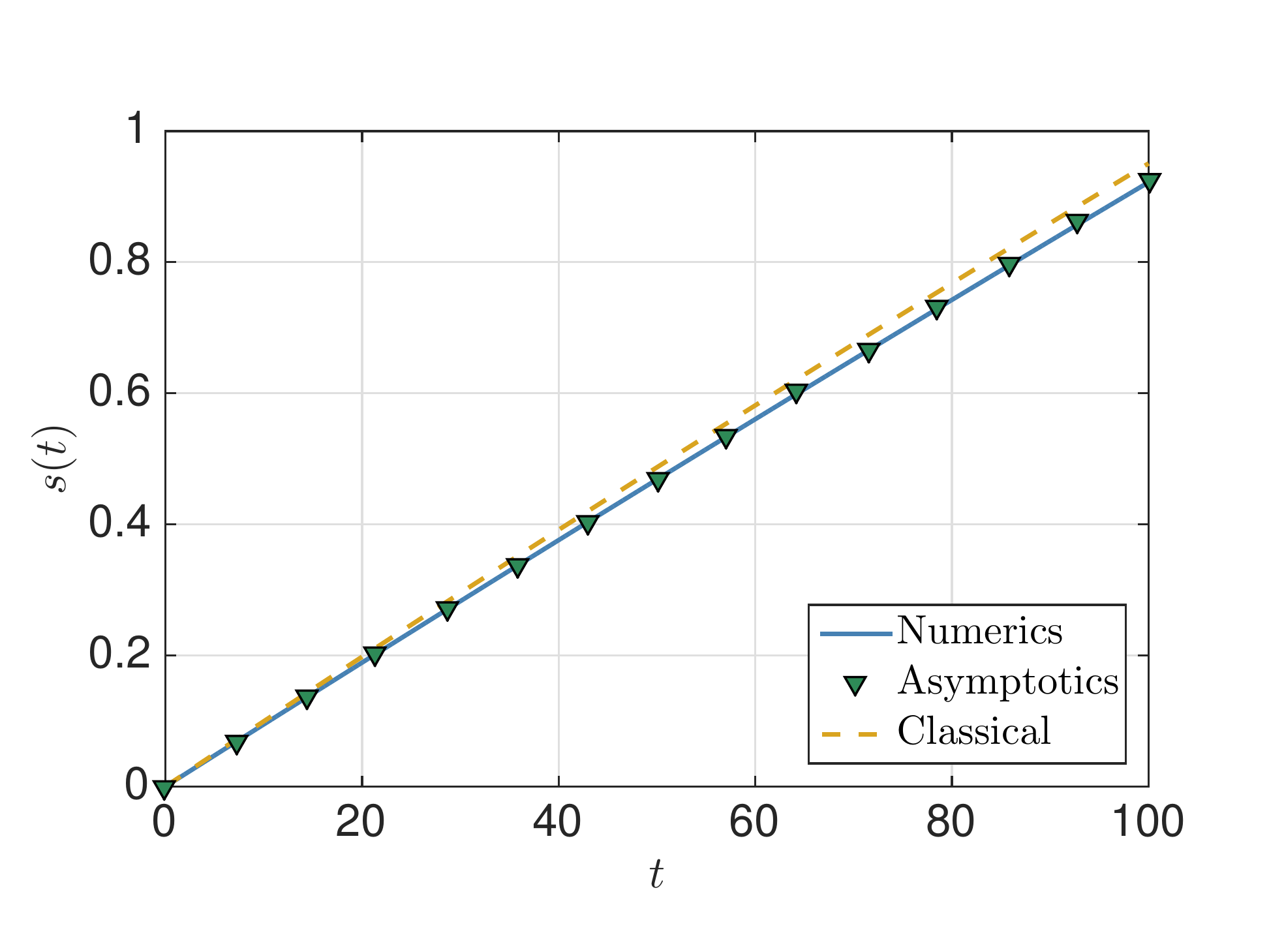}
    \caption{$\beta=10$, $\Bi=0.1$.}\label{fig:results_b}
	\end{subfigure}
	
	\begin{subfigure}{.47\textwidth}
    \centering
    \includegraphics[width=\textwidth]{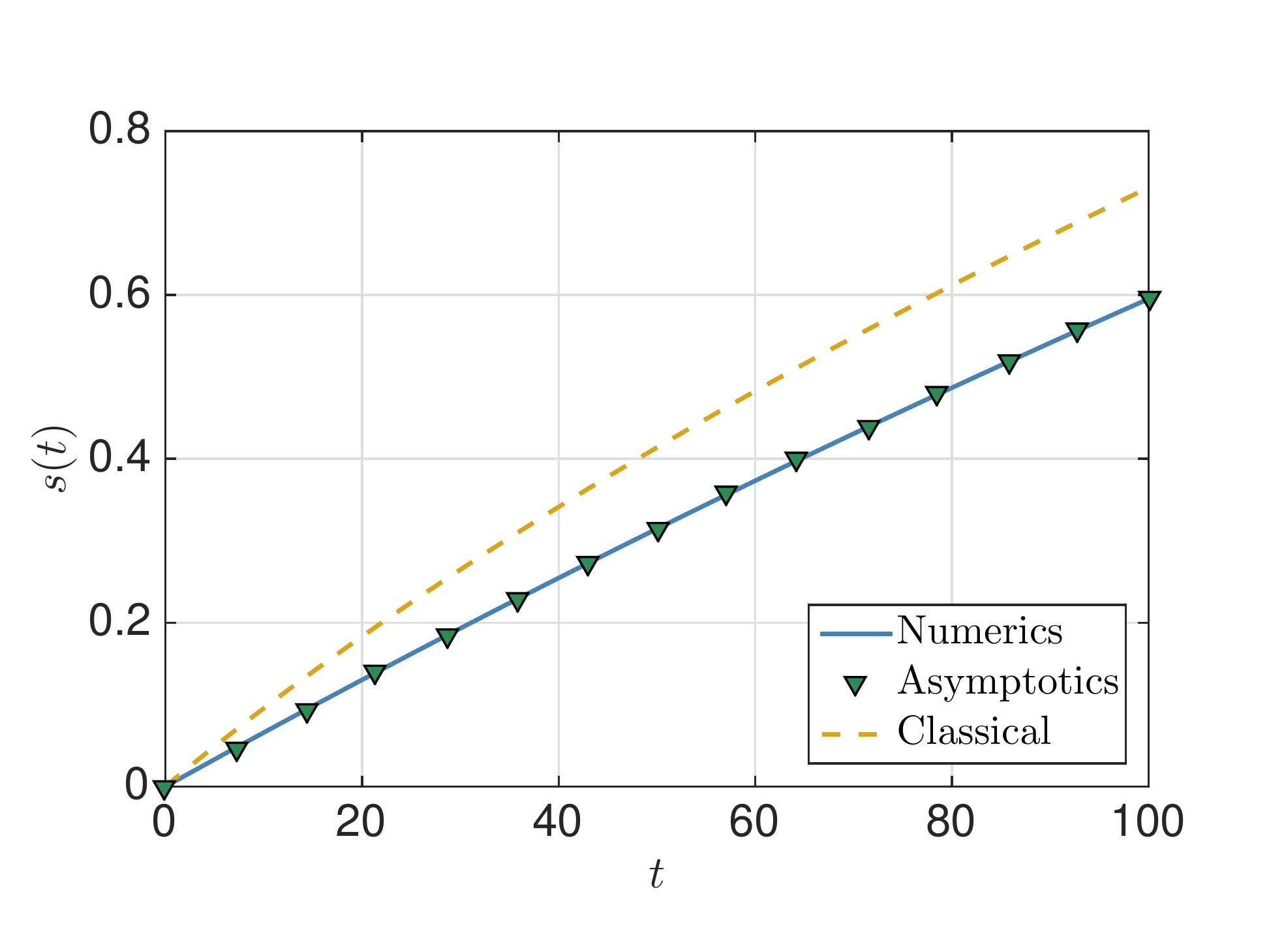}
    \caption{$\beta=100$, $\Bi=1$.}\label{fig:results_c}
	\end{subfigure}
	~
	\begin{subfigure}{.47\textwidth}
    \centering
    \includegraphics[width=\textwidth]{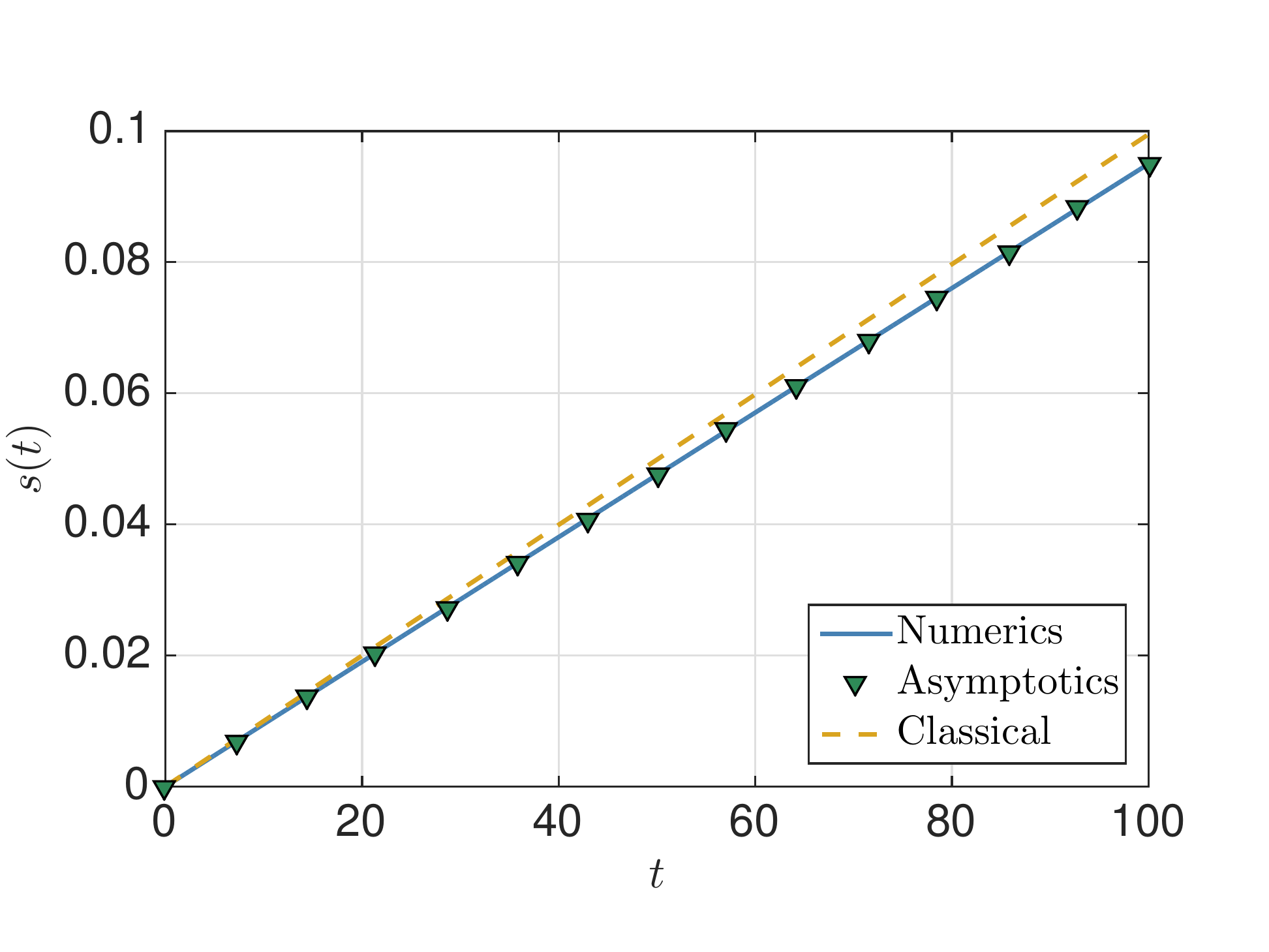}
    \caption{$\beta=100$, $\Bi=0.1$.}\label{fig:results_d}
	\end{subfigure}
	\caption{Evolution of the solid-liquid interface for $\Bi=1$ and $\beta=1,10$. Solid lines correspond to the numerical solutions, symbols refer to the solutions of Eq. \eqref{asy:finalODE}. The dashed line shows the corresponding classical solution.}\label{fig:results}
\end{figure}

In Fig. \ref{fig:results} we show the numerical and asymptotic solutions of the non-classical formulation and compare them against the classical solution for constant thermal conductivity. This is done for different choices of $\Bi$ and $\beta$.

Firstly, we observe that the asymptotic solution is in good agreement with the numerical solution. Only in Fig. \ref{fig:results_a} do some discrepancies appear for larger times. This is due to the ratio $\Bi/\beta$, whose value in the case of Fig. \ref{fig:results_a} is at least one order of magnitude larger than in the other cases. Recall that the asymptotic solution is based on the assumption $\Bi/\beta\ll1$ and hence we should account for higher-order terms to increase the accuracy of the asymptotic solution.

Secondly, we observe that the speed of the solid-liquid interface decreases as we decrease the Biot number and the usual square-root profile transforms into a linear one. Furthermore, as we decrease Bi both classical and non-classical formulations tend to the same solution and non-local effects described by the size-dependent ETC seem to become less important. For $\Bi=0.01$, which we do not show here, both the classical and non-classical profiles are indistinguishable for $\beta\in(10,100)$. In fact, our asymptotic analysis indicates that we expect a linear profile until $t=O(\beta/\Bi)$ with a slope $\Bi/\beta$. The same slope is obtained in the small-time analysis of the classical formulation (see \ref{appendix}). Physically, the convergence of both solutions to the same profile can be understood by recalling that small Biot numbers correspond to a situation where heat flow is limited by the heat exchange with the environment rather than heat conduction through the bulk \cite{Incropera2013}, which implies that the form of thermal conductivity employed should not matter in this case. 

In Fig. \ref{fig:results_c} it can also be observed that large deviations between the non-classical and classical formulations appear for large values of $\beta$. This may be understood by recalling that the Stefan number corresponds to a slow solidification, since $s_t\propto\beta^{-1}$. Hence, non-local effects remain for a longer time period due to the small growth rate, which causes these larger deviations between both formulations.

\begin{figure}[h!]
	\centering
	\begin{subfigure}{.47\textwidth}
    \centering
    \includegraphics[width=\textwidth]{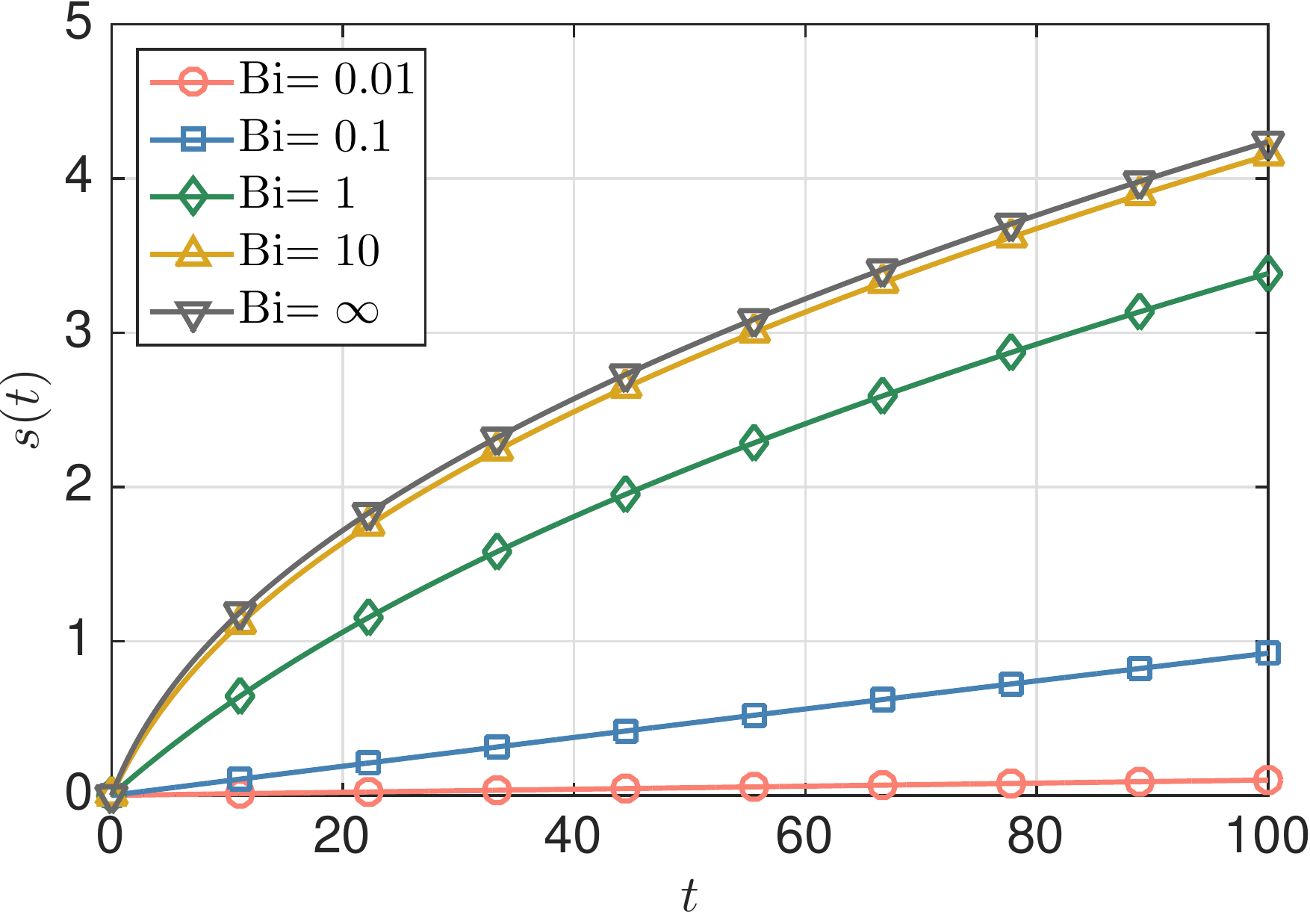}
    \caption{Effective Fourier law, $\beta=10$.}\label{fig:results2_a}
	\end{subfigure}
	~
	\begin{subfigure}{.47\textwidth}
    \centering
    \includegraphics[width=\textwidth]{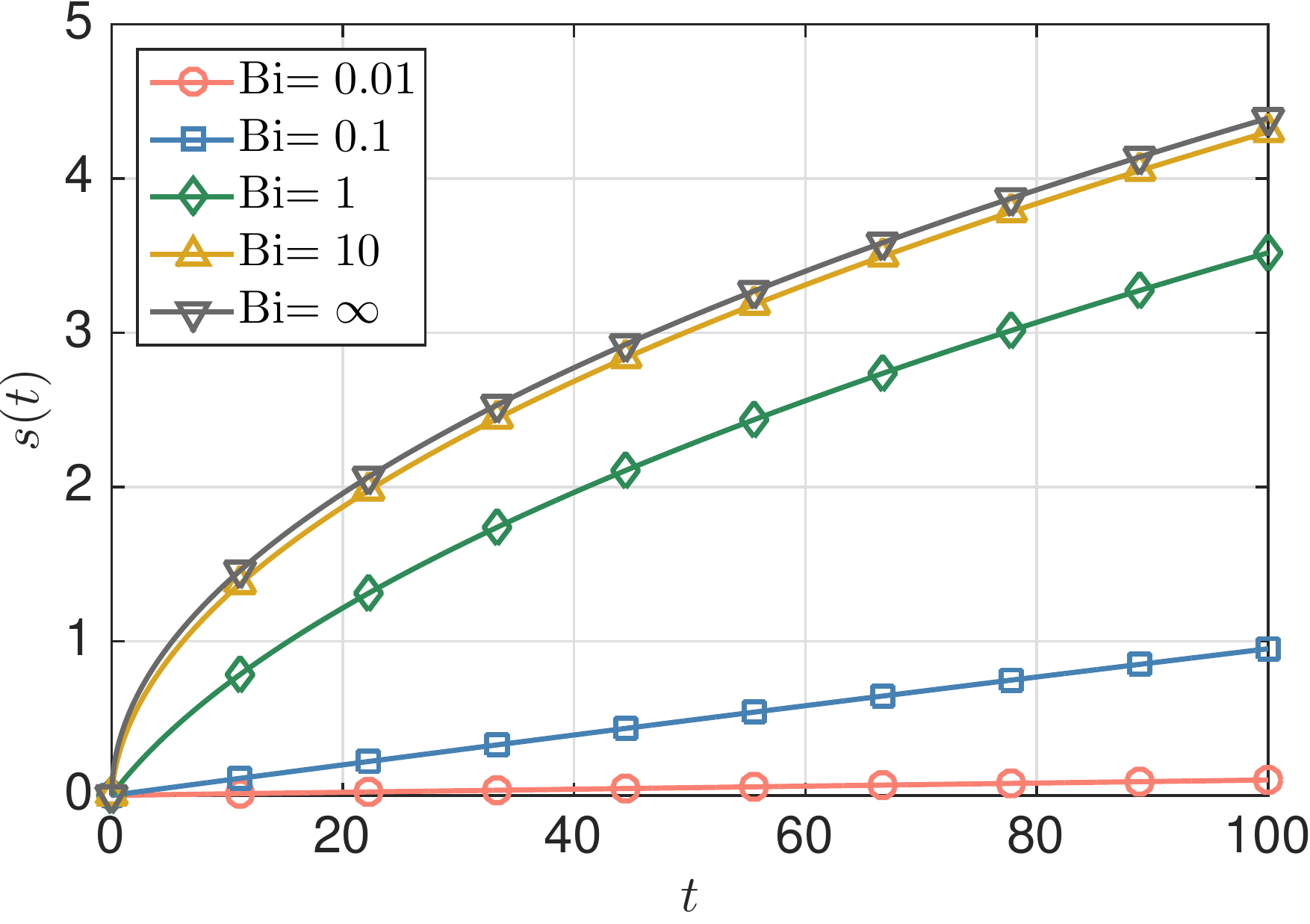}
    \caption{Classical Fourier law, $\beta=10$.}\label{fig:results2_b}
	\end{subfigure}
	
	\begin{subfigure}{.47\textwidth}
    \centering
    \includegraphics[width=\textwidth]{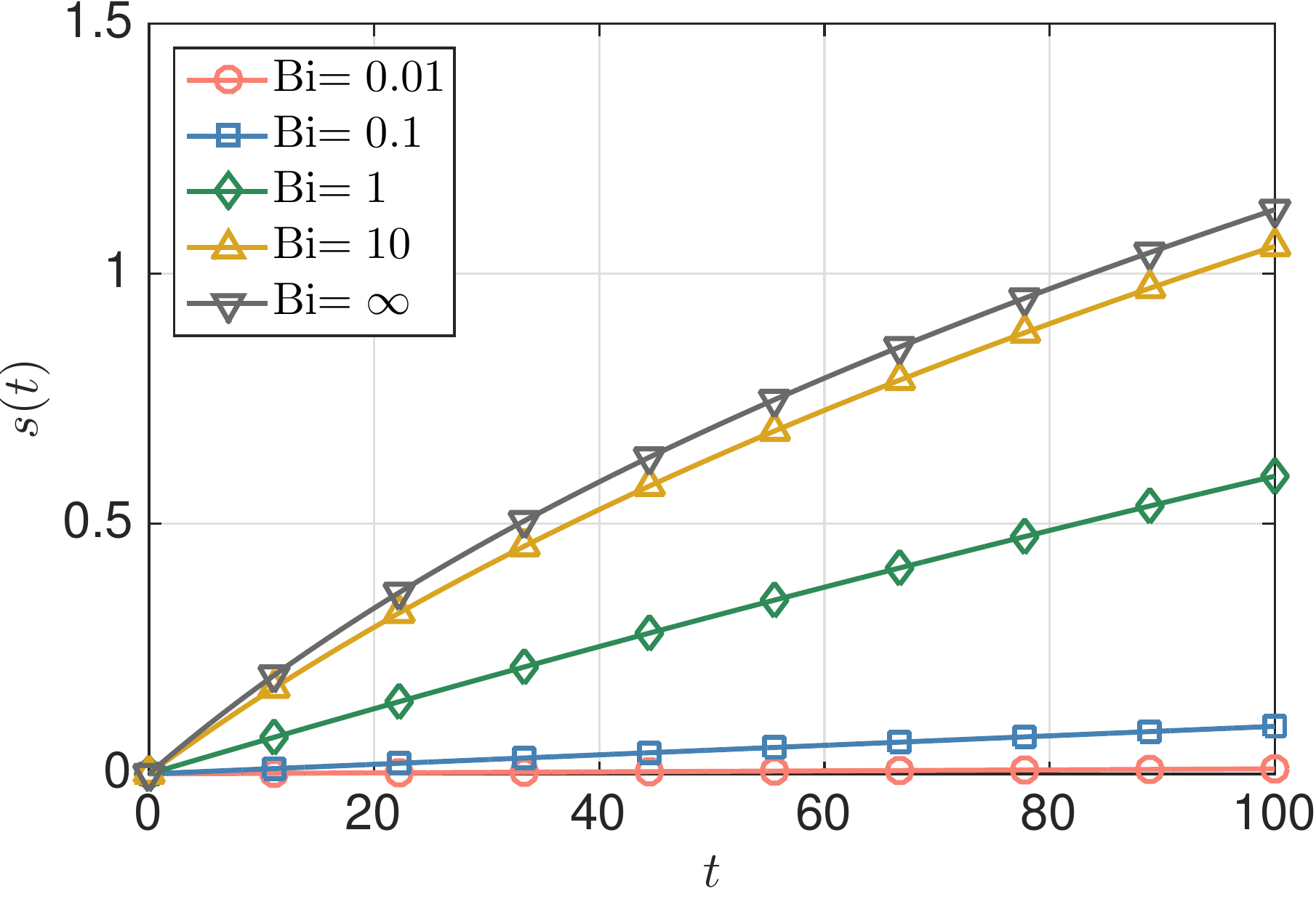}
    \caption{Effective Fourier law, $\beta=100$.}\label{fig:results2_c}
	\end{subfigure}
	~
	\begin{subfigure}{.47\textwidth}
    \centering
    \includegraphics[width=\textwidth]{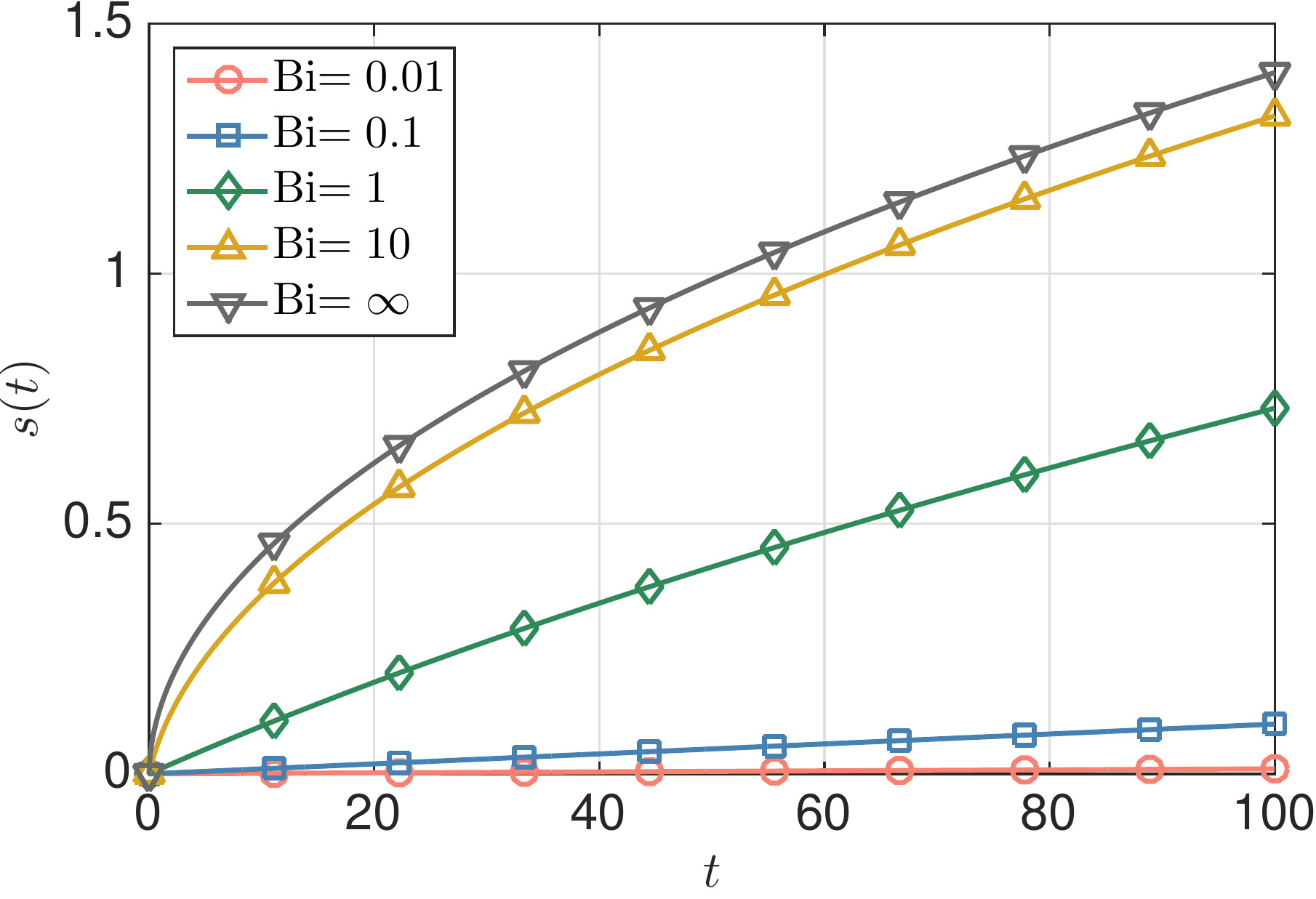}
    \caption{Classical Fourier law, $\beta=100$.}\label{fig:results2_d}
	\end{subfigure}
	\caption{Evolution of the solid-liquid interface for different values of Bi, according to the non-classical (panels (a) and (c)) and classical (panels (b) and (d)) formulations. The case $\Bi=\infty$ corresponds to the fixed-temperature condition $T(0,t)=-1$.}\label{fig:results2}
\end{figure}

\begin{figure}[h!]
	\centering	
	\begin{subfigure}{.47\textwidth}
    \centering
    \includegraphics[width=\textwidth]{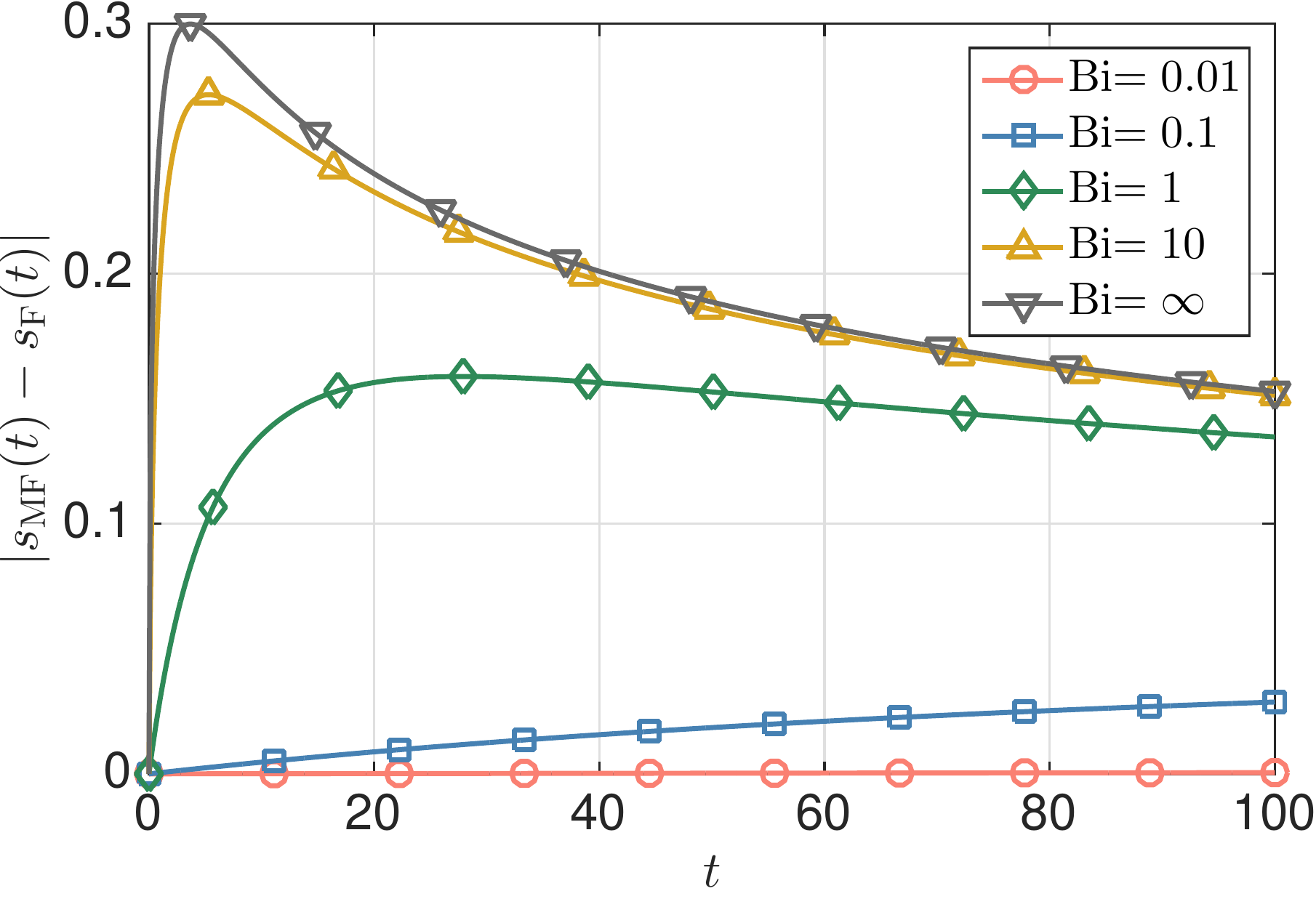}
    \caption{$\beta=10$.}\label{fig:results3_a}
	\end{subfigure}
	~
	\begin{subfigure}{.47\textwidth}
    \centering
    \includegraphics[width=\textwidth]{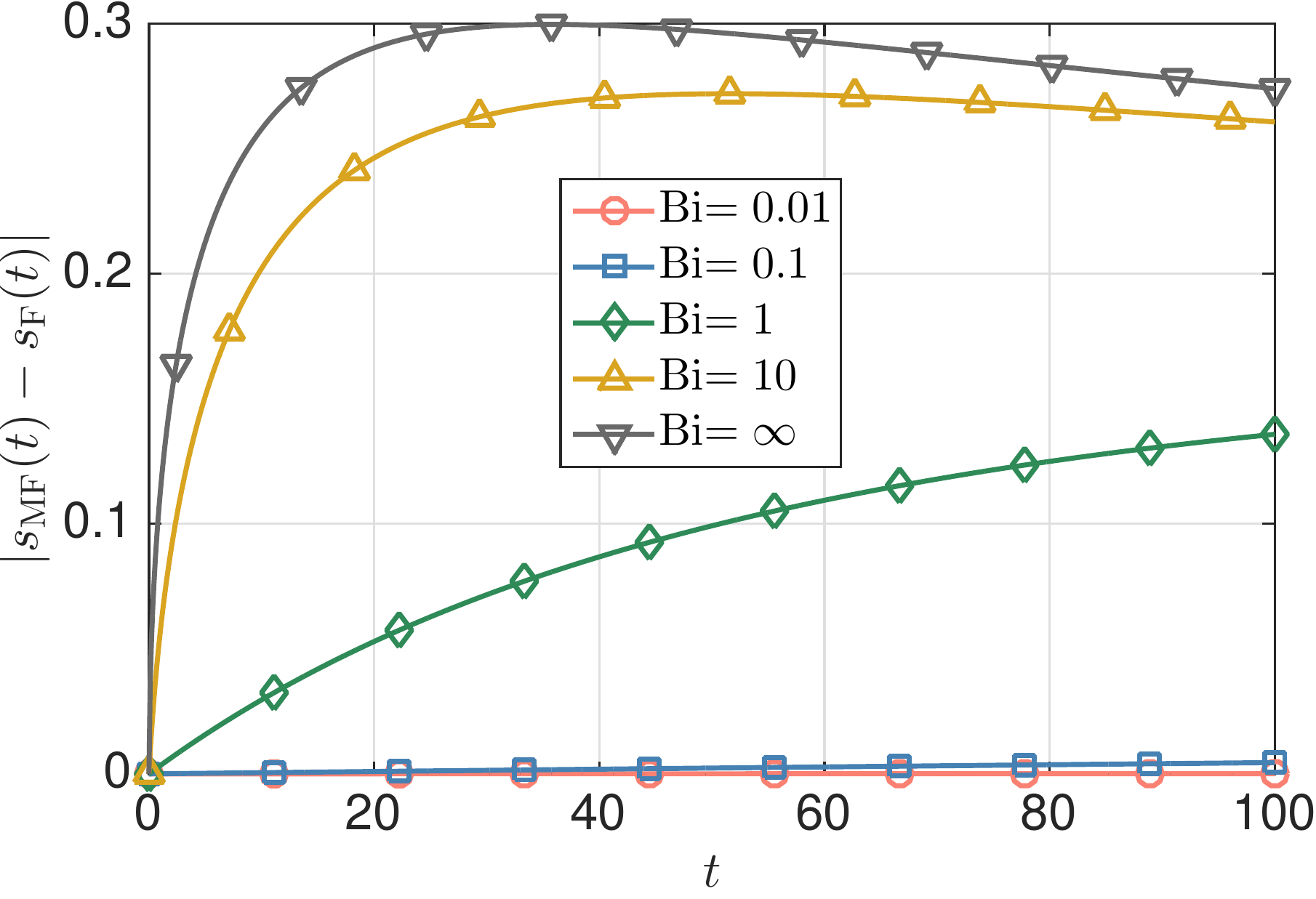}
    \caption{$\beta=100$.}\label{fig:results3_b}
	\end{subfigure}
	\caption{Evolution of the absolute difference $|s_\text{NF}(t)-s_\text{F}(t)|$, where $s_\text{MF}$ and $s_\text{F}$ represent the position of the interface according to the modified and classical formulations.}\label{fig:results3}
\end{figure}

To illustrate the general case of an arbitrary Biot number, in Fig. \ref{fig:results2} we show the evolution of the solid-liquid interface for different values of the Stefan and Biot numbers according to the classical and non-classical formulations. For $\beta=10$, differences between both formulations are considerably less than for the larger value $\beta=100$. In Fig. \ref{fig:results3} we have plotted the evolution in time of the absolute difference of both formulations for different values of $\beta$ and Bi. We observe the existence of two regimes: in the first regime, due to the presence of non-local effects, differences between the classical and modified formulations increase, whereas in the second time regime they decrease as non-local effects disappear. By observing Figs. \ref{fig:results2_a} we see that this change in behaviour occurs when $s\approx1$. In the case of $\beta=100$ the second time regime enters later due to the fact that solidification is slower and hence the presence of non-local effects is significant for a longer period of time.

Hence, these results suggest that non-local effects become less important if we decrease the Biot number, which corresponds to a poor heat conduction through the solid, or by decreasing the Stefan number, which corresponds to fast solidification. In the first case, the choice of thermal conductivity does not alter the evolution of $s(t)$ significantly. In the latter, non-local effects disappear due to the fast growth of the solid phase.

\begin{figure}[h!]
	\centering
	\begin{subfigure}{.47\textwidth}
    \centering
    \includegraphics[width=\textwidth]{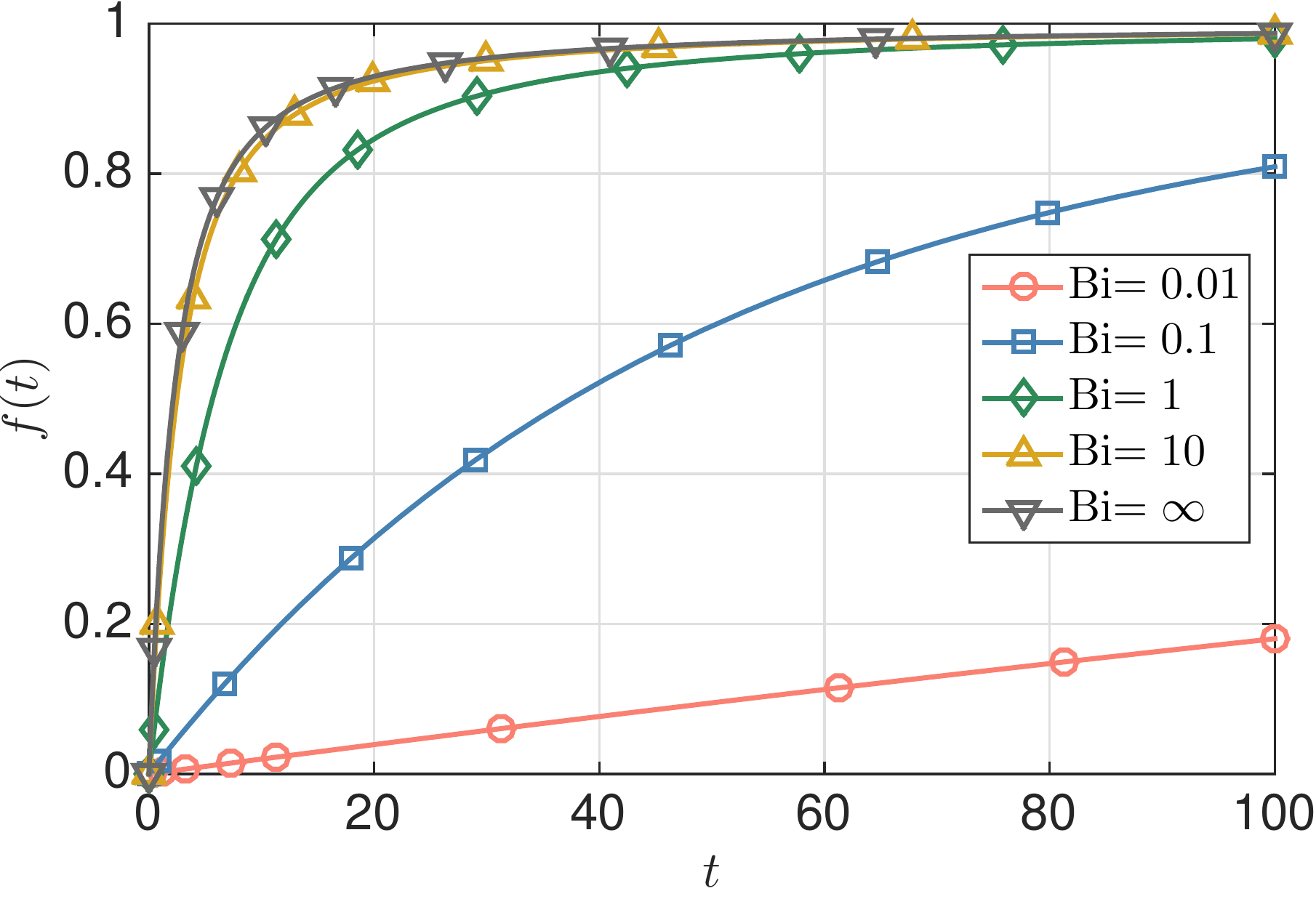}
   \caption{$\beta=10$.}
	\end{subfigure}
	~
	\begin{subfigure}{.47\textwidth}
    \centering
    \includegraphics[width=\textwidth]{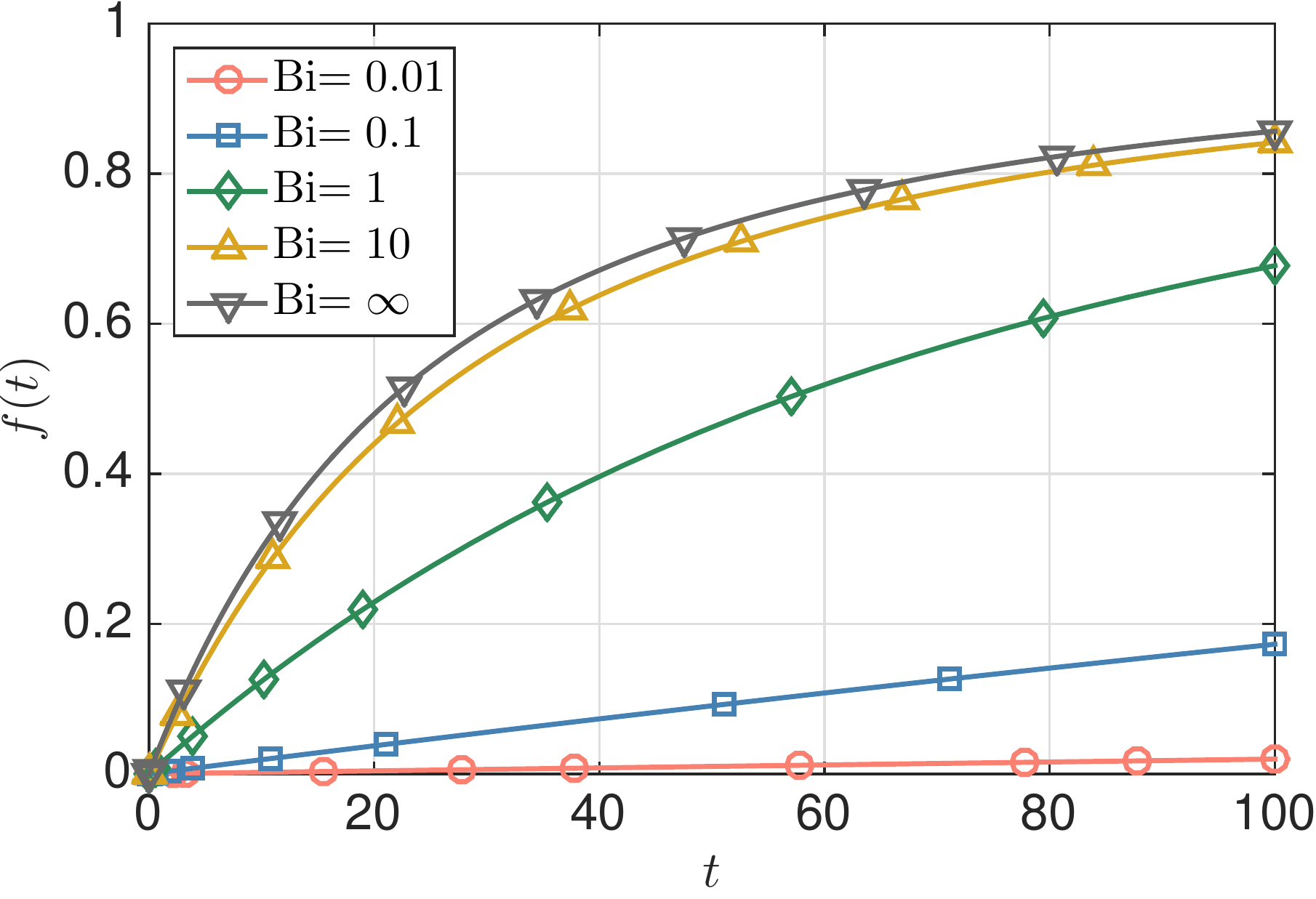}
    \caption{$\beta=100$.}
	\end{subfigure}
	\caption{Evolution of the effective thermal conductivity as a function of time (computed as $f(t)=f(s(t))$) for different values of the Biot number Bi. The case $\Bi=\infty$ corresponds to the fixed temperature condition $T(0,t)=-1$.}\label{fig:conductivity}
\end{figure}

\begin{figure}
	\centering
	\begin{subfigure}{.47\textwidth}
    \centering
    \includegraphics[width=\textwidth]{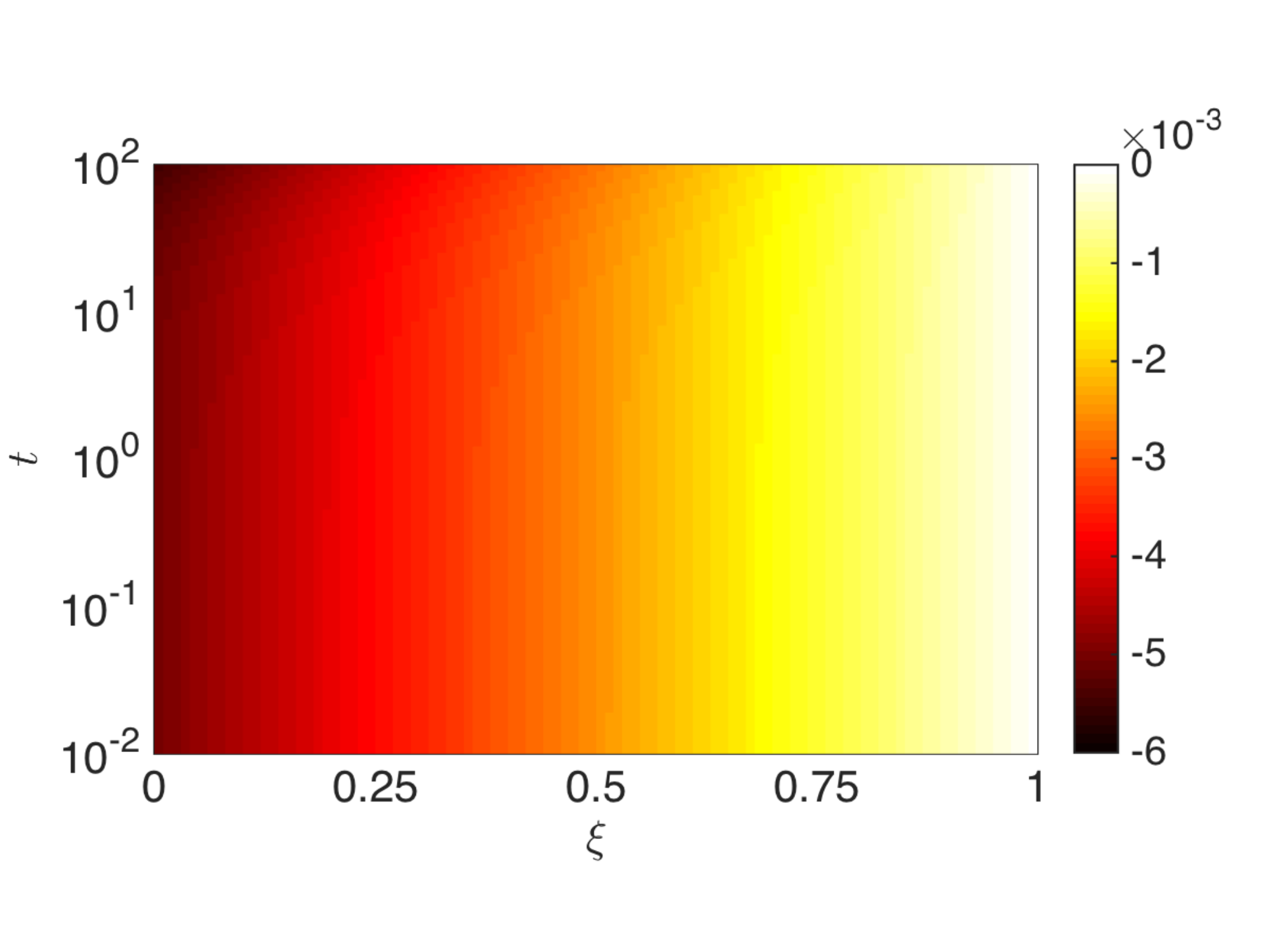}
    \caption{Effective Fourier law, $\Bi=0.01$.}\label{fig:resultsTemp_a}
	\end{subfigure}
	~
	\begin{subfigure}{.47\textwidth}
    \centering
    \includegraphics[width=\textwidth]{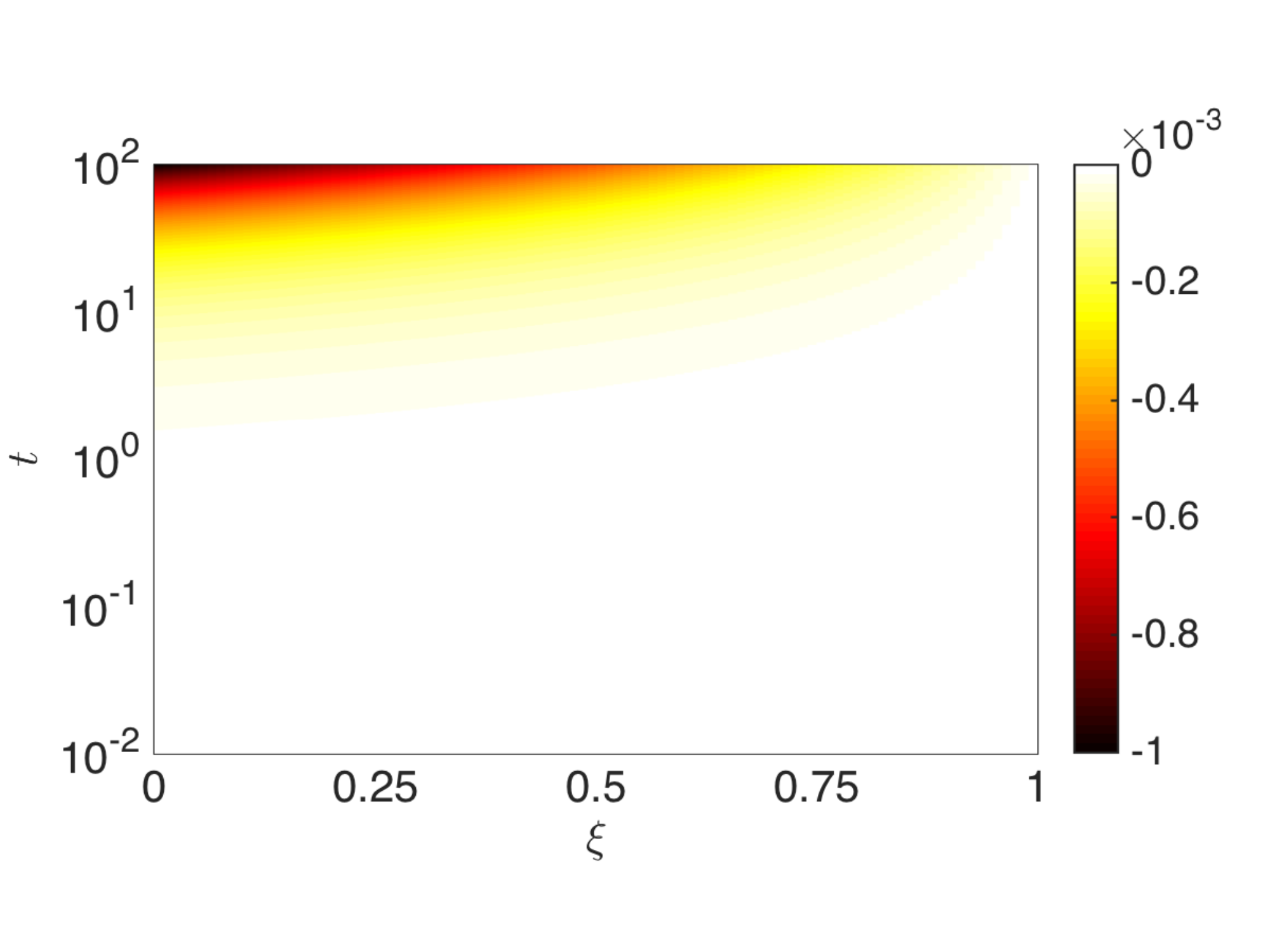}
    \caption{Classical Fourier law, $\Bi=0.01$.}\label{fig:resultsTemp_b}
	\end{subfigure}
	
	\begin{subfigure}{.47\textwidth}
    \centering
    \includegraphics[width=\textwidth]{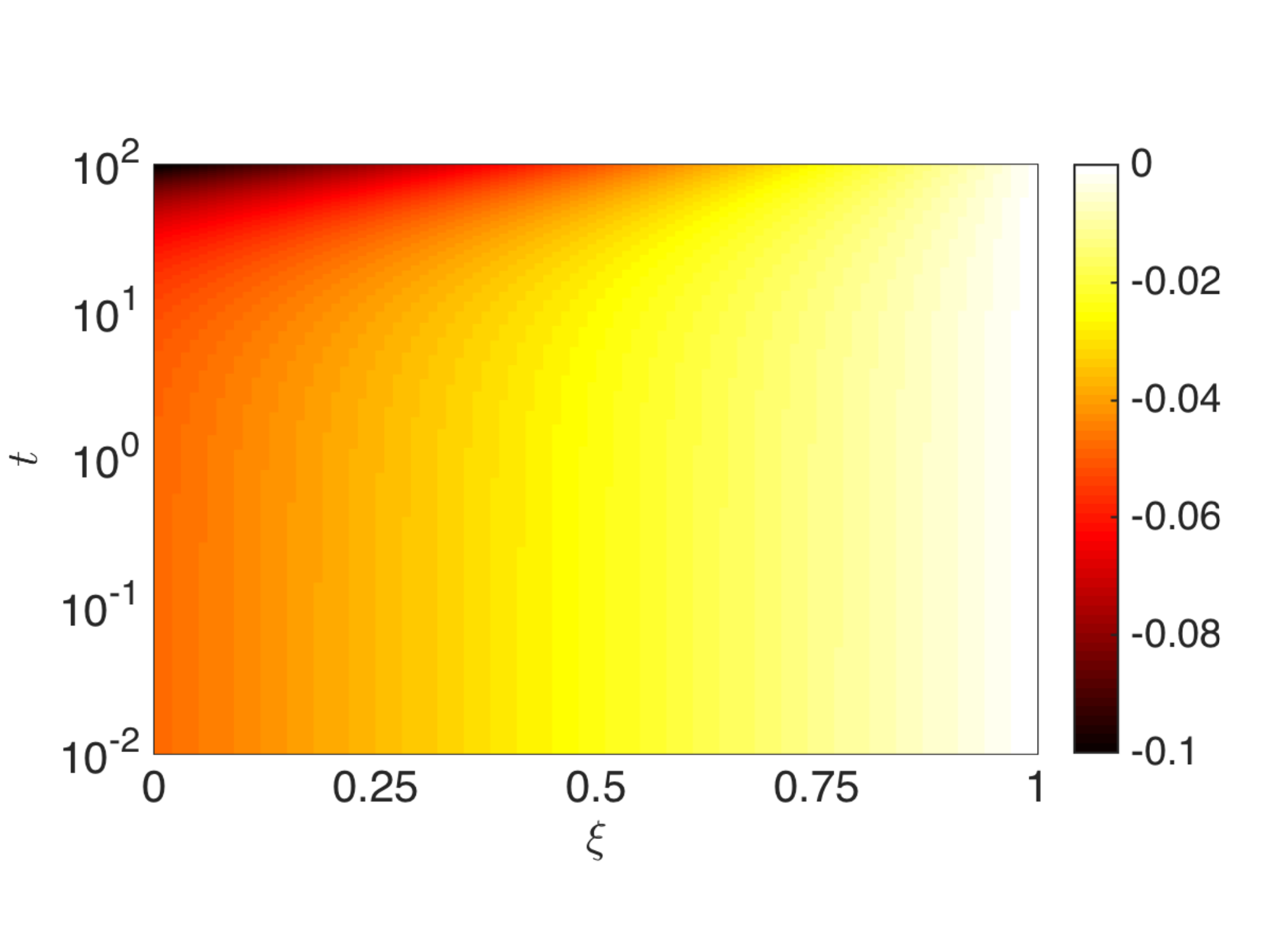}
    \caption{Effective Fourier law, $\Bi=0.1$.}\label{fig:resultsTemp_c}
	\end{subfigure}
	~
	\begin{subfigure}{.47\textwidth}
    \centering
    \includegraphics[width=\textwidth]{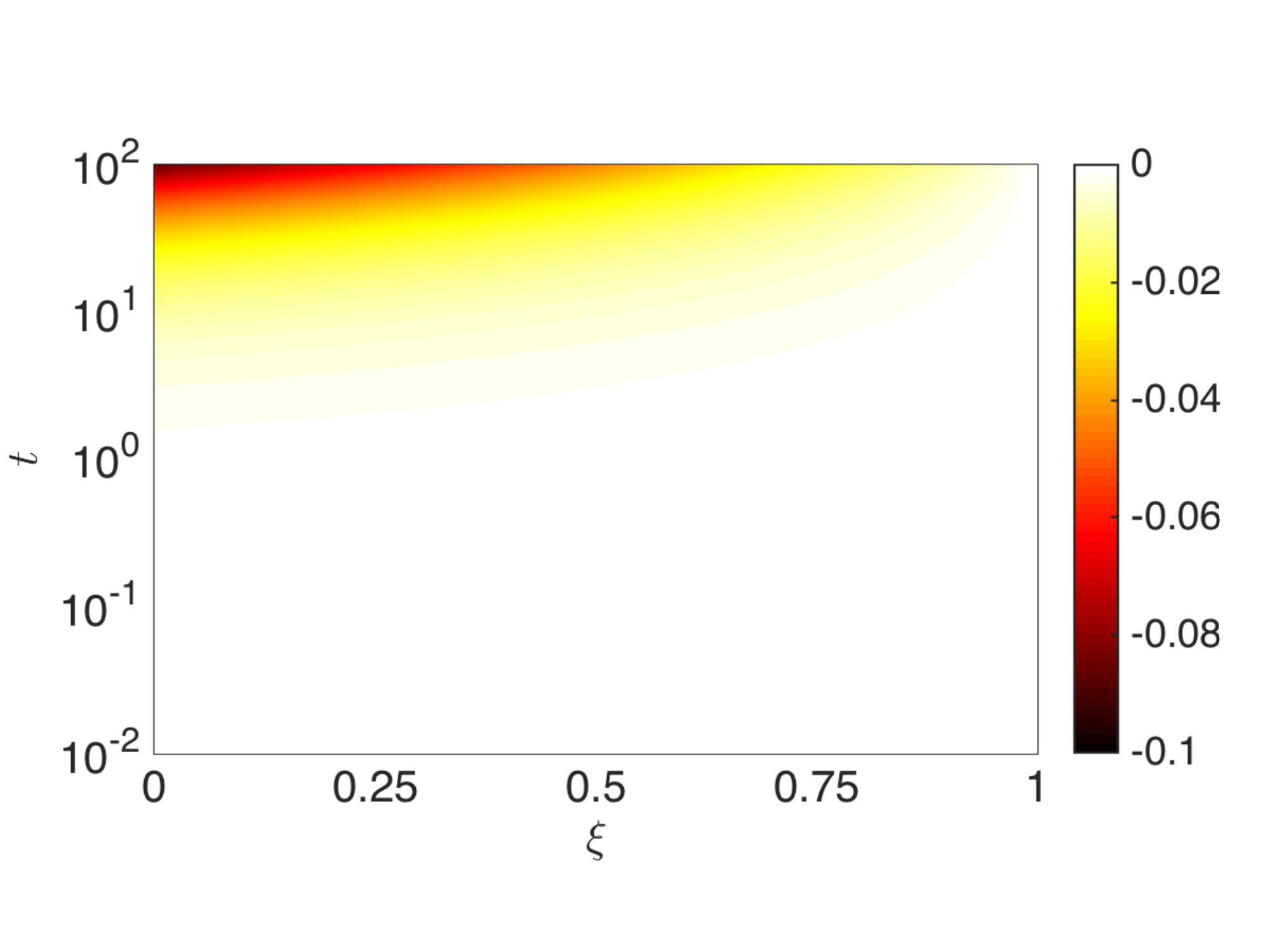}
    \caption{Classical Fourier law, $\Bi=0.1$.}\label{fig:resultsTemp_d}
	\end{subfigure}
	
	\begin{subfigure}{.47\textwidth}
    \centering
    \includegraphics[width=\textwidth]{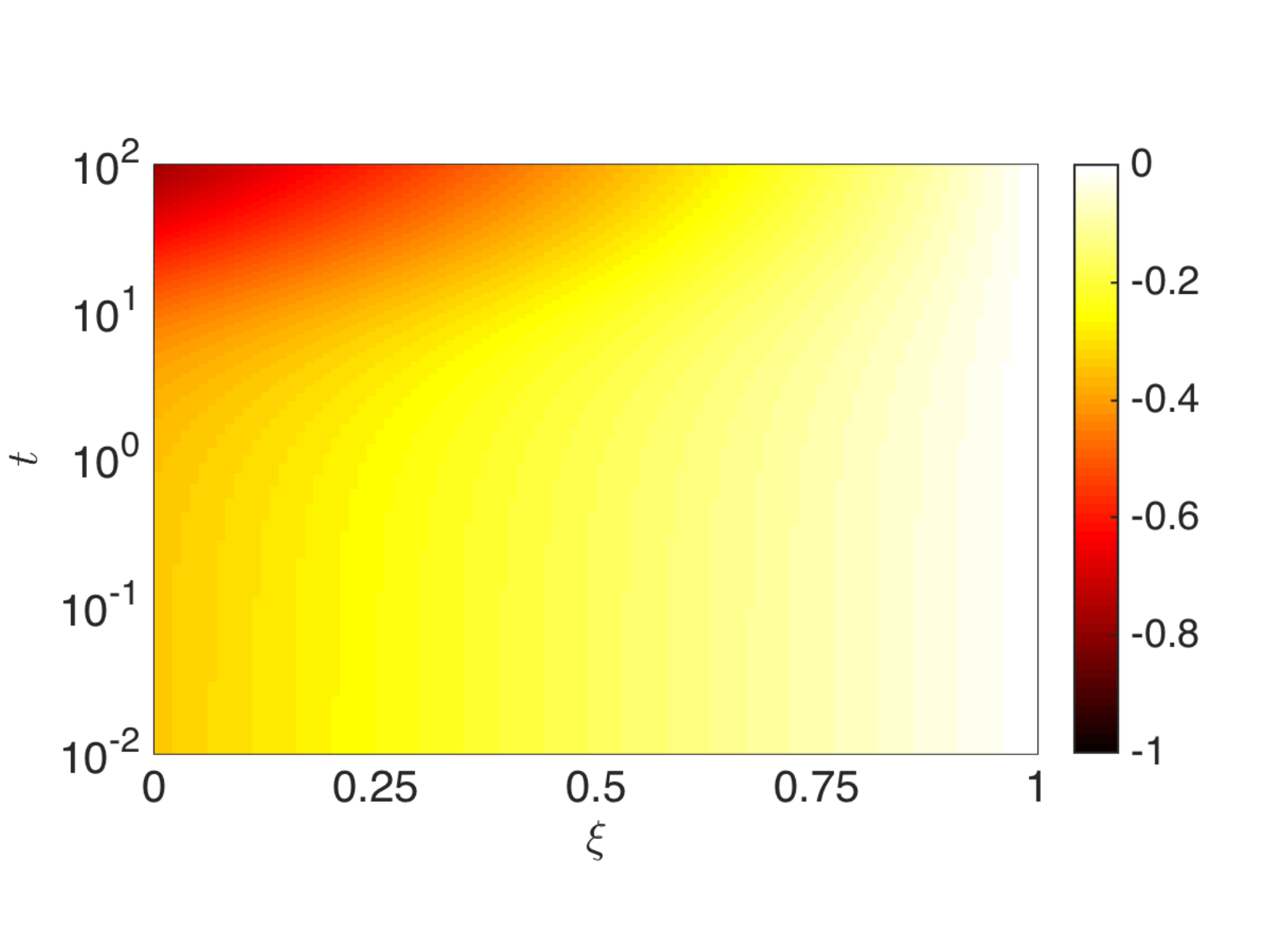}
    \caption{Effective Fourier law, $\Bi=1$.}\label{fig:resultsTemp_e}
	\end{subfigure}
	~
	\begin{subfigure}{.47\textwidth}
    \centering
    \includegraphics[width=\textwidth]{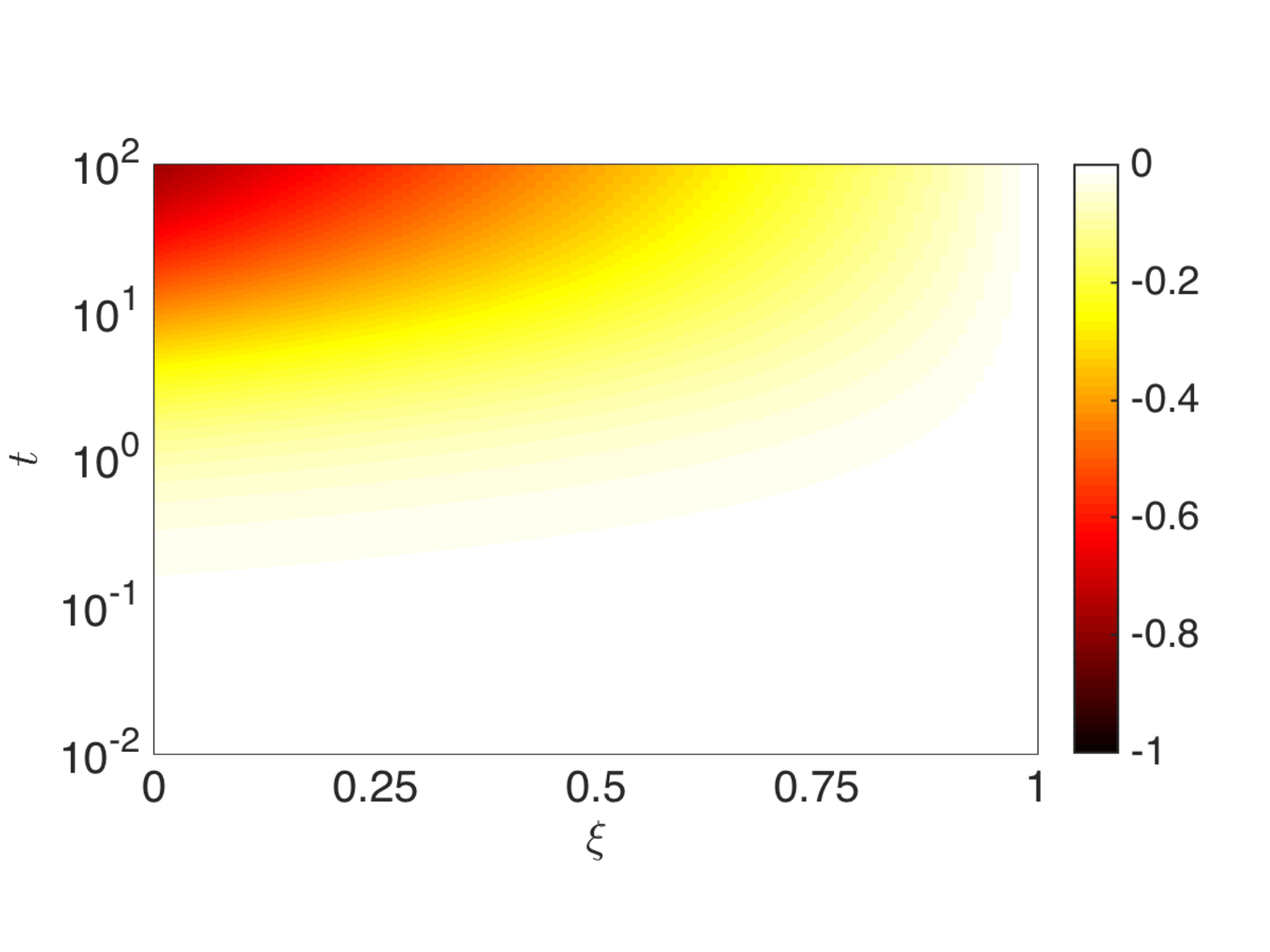}
    \caption{Classical Fourier law, $\Bi=1$.}\label{fig:resultsTemp_f}
	\end{subfigure}
	\caption{Heat maps showing the evolution of the temperature according to the effective and classical Fourier laws in the transformed coordinate $\xi$. In all the panels the value $\beta=10$ has been used.}
	\label{fig:resultsTemp}
\end{figure}

The effect of the Biot number on the temperature can be observed in Fig. \ref{fig:resultsTemp}, which shows the evolution of the temperature profiles according the non-classical and classical formulations for different values of Bi and for $\beta=10$. For early times the size of the solid is small, which leads to a small ETC and hence changes in temperature in the non-classical formulation occur mainly near $x=0$. This happens independent of the value of Bi, as it can be seen by comparing panels (a), (c) and (e). In addition, it can seem that diffusion is enhanced as Bi decreases, in contradiction to the fact that conduction becomes worse in this limit. We must recall that the spatial coordinate used in Fig. \ref{fig:resultsTemp} is the alternative variable $\xi=x/s$, and hence the growth rate of the solid is not shown here. Hence, the temperature profiles must be observed together with the growth of the solid for these values of Bi, shown in Figs. \ref{fig:results2_a}. Another indicator for a worse conduction for small values of Bi is the range of temperatures shown in the color bars of the corresponding plots.\\
In the classical formulation, diffusion through the solid occurs normally and changes in temperature appear gradually through the crystal. Similarly as in the case of the non-classical formulation, panels (b), (d) and (f) of Fig. \ref{fig:resultsTemp} must be observed by having in mind the corresponding growth rates of the solid-liquid in, shown in Fig. \ref{fig:results2_b}.\\
We can also see that the initial differences between the temperature profiles for a fixed Bi disappear as time increases, which is caused by the fact that the growth of the solid leads to an increase of the ETC, which finally converges to the classical conductivity.

\section{Conclusions}\label{sec:conclusion}

In this paper we have formulated a mathematical model for a one-dimensional solidification process with a size-dependent thermal conductivity and a Newton cooling condition at the fixed boundary. Numerical and asymptotic solution methods have been proposed and compared, showing good agreement for the range of parameters considered. In contrast to previous studies, we have studied the limit of a small Biot number, which seems more reasonable after a proper analysis. We have observed that non-local effects tend to disappear as the Biot number decreases, which seems surprising since the initial solidification rate is proportional to Bi and thus non-local effects would have been expected to keep present for longer periods of time. This result has been attributed to the poor conduction in both the classical and the modified formulations due to the Biot number being small, since this implies that the heat exchange with the environment at the fixed boundary dominates over the heat conduction through the growing solid. Similarly, as we decrease the Stefan number the presence of non-local effects is reduced to an initial stage of the process that is overcome rapidly due to the higher speed of the solid-liquid interface. 


\section*{Acknowledgements}
M.~C. acknowledges that the research leading to these results has been funded by ``La Caixa" Foundation and by the CERCA programme of the Generalitat de Catalunya. The author thanks T.~G. Myers and M.~G. Hennessy for a critical reading of the manuscript during its preparation.

\appendix

\section{Small-time behaviour of the classical Stefan problem with Newton cooling conditions}
\label{appendix}

As in the small-time analysis for the non-classical case, we assume $\Bi,\beta=O(1)$. After applying the transformation $T(x,t)=u(\xi,t)$, the classical problem with Newton conditions takes the form
\begin{subequations}\label{classical:eqsfixed}
\begin{alignat}{3}
	&s^2u_t=\xi ss_tu_\xi+u_{\xi\xi},\qquad &&0\leq \xi\leq 1,\label{classical:heat}\\
	&u_\xi=s\Bi(1+u),\qquad &&\xi=0,\label{classical:bc0}\\
	&u=0,\qquad &&\xi=1,\label{classical:bc1}\\
	&\beta ss_t=u_\xi,\qquad &&\xi=1,\label{classical:Stefan}
\end{alignat}
\end{subequations}
Let $t\ll1$. Assuming only $s=O(t^p)$ requires $p=0.5$ to balance terms in the Stefan condition, which is unphysical since we recover the initially infinite solidification rate which should be avoided due to the Newton condition. Furthermore, only scaling $s$ transforms the Newton condition into $u_\xi\approx0$ and therefore no heat exchange would occur at the boundary to drive the solidification process. In fact, balance between both sides of Eq. \eqref{classical:bc0} can only happen if $u=O(t^p)$. Assuming $u,s=O(t^p)$ in Eq. \eqref{classical:Stefan} implies that $p=1$ is required for balancing terms. Hence, for an arbitrary $\varepsilon\ll1$ we define the scaled variables $u=\varepsilon\bar u$, $s=\varepsilon\bar s$, $t=\varepsilon\bar t$ and $\xi=\bar \xi$. 
The Stefan condition becomes
\begin{equation}\label{classical:Stefan2}
	\beta\bar s\bar s_{\bar t}=\bar u_{\bar \xi},
\end{equation}
where $\bar u_{\bar\xi}$ is evaluated at $\bar\xi=1$. Upon neglecting terms of order $\varepsilon$, the problem for $\bar u$ is
\begin{subequations}\label{classical:ODEu}
\begin{align}
	&\bar u_{\bar \xi\bar \xi}=0,\\
	&\bar u_{\bar \xi}(0)=\bar s\Bi,\\
	&\bar u(1)=0,
\end{align}
\end{subequations}
which gives $\bar u(\bar\xi)=\bar s\Bi(\bar\xi-1)$. Substituting this expression into Eq. \eqref{classical:Stefan2} yields the initial value problem
\begin{equation}
	\beta \bar s\bar s_{\bar t}=\Bi,\qquad \bar s(0)=0,
\end{equation}
whose solution is $\bar s(\bar t)=\Bi\beta^{-1} \bar t$. In the original dimensionless variables, the small-time behaviour of the classical problem is therefore
\begin{equation}
	u(\xi,t)=\Bi^2\beta^{-1}t(\xi-1),\qquad s(t)=\Bi\beta^{-1} t.
\end{equation}

\section*{References}
\bibliographystyle{elsarticle-num}
\bibliography{References}

\end{document}